\documentclass[reprint, amsmath,amssymb,
 aps,nofootinbib]{revtex4-2}
\usepackage{color}    
\usepackage{graphicx}
\usepackage{dcolumn}
\usepackage{bm}
\usepackage{mathrsfs}
\usepackage{bm}
\usepackage{amsmath}
\usepackage{amssymb}
\usepackage{mathtools}
\usepackage{tocloft}
\usepackage{url}
\usepackage{physics}
\usepackage{float}
\usepackage{amsthm}  
\usepackage{stmaryrd, bbm, slashed, simplewick, bbold} 
\usepackage[hidelinks]{hyperref}
\usepackage{cleveref}
\usepackage{csquotes}
\usepackage{caption}
\usepackage{setspace}
\DeclareGraphicsExtensions{.eps}
\usepackage{tikz}
\colorlet{RED}{red}

\newcommand{\caldist}[2]{\Delta (#1,#2)}

\makeatletter
\newcommand{\fmarki}{*}
\newcommand{\fmarkii}{\ensuremath{\dagger}}
\newcommand{\fmarkiii}{\ensuremath{\ddagger}}
\newcommand{\fmarkiiii}{\ensuremath{$\mathsection$}}
\def\@fnsymbol#1{{\ifcase#1\or \fmarki\or \fmarkii\or \fmarkiii\or \fmarkiiii \else\@ctrerr\fi}}
\makeatother
\renewcommand{\fmarki}{*}
\renewcommand{\fmarkii}{$\flat$}
\renewcommand{\fmarkiii}{$\mathsection$}
\renewcommand{\fmarkiiii}{$\natural$}

\begin{document}

\preprint{APS/123-QED}

\title{Curvature from multipartite entanglement in quantum gravity states}
\author{Simone Cepollaro}
\email{simone.cepollaro-ssm@unina.it}
\affiliation{Scuola Superiore Meridionale, Largo S. Marcellino 10, 80138 Napoli, Italy; INFN, Sezione di Napoli, Italy;} 
\author{Goffredo Chirco}
\email{goffredo.chirco@unina.it}
\author{Gianluca Cuffaro}
\email{g.cuffaro@studenti.unina.it}
\author{Vittorio D'Esposito}
\email{vittorio.desposito@unina.it}
\affiliation{Dipartimento di Fisica ``Ettore Pancini'', Universit\`a di Napoli Federico II, Napoli, Italy; INFN, Sezione di Napoli, Italy.}

\date{\today}

\begin{abstract}

We investigate the multipartite entanglement of a uniformly curved quantum $3D$ space region with boundary, realised in terms of spin networks defined on a graph with non trivial $SU(2)$ holonomies, in the framework of loop quantum gravity. The presence of intrinsic curvature in the region is encoded in closure (topological) defects associated with \emph{tag}-spins attached to the vertices of the graph.
For such states, we generalise the bulk-to-boundary mapping as to include the space of tags in an extended boundary space: bulk information is \emph{shared} among generically entangled boundary surfaces and intrinsic curvature degrees of freedom. We model the quantum region on a tripartite system composed by two (complementary) boundary subregions and the set of bulk tags. 
Via replica techniques, we can compute the typical value of the logarithmic negativity of the reduced boundary, described as an open quantum system, in a large spin regime. 

We find three entanglement regimes, depending on the ratio between the number of tags (curvature) and the area of the dual surface at the boundary. These are well described by the \emph{generalised Page curve} of a tripartite random state. In particular, we find area scaling behaviour for negativity in case of small curvature, while for large curvature the negativity vanishes, suggesting an effective thermalization of the boundary. Remarkably, the PPT character of the mixed boundary state corresponds to a change in the effective topology of the network, with the two boundary subregions becoming disconnected.

\begin{description}
\item[Keywords] Quantum gravity, random tensor networks, multipartite entanglement, logarithmic negativity

\end{description}
\end{abstract}
\maketitle
\section{Introduction} 

In the last two decades, the idea that continuum spacetime geometry may be emerging from entanglement has radically changed all approaches to quantum gravity \cite{VanRaamsdonk:2010pw, Swingle:2009bg, Bianchi_2014, Lashkari:2013koa, Faulkner_2014, Jacobson_2016, PhysRevD.95.024031, Swingle:2014uza, Qi:2013caa}. Beside the AdS/CFT framework~\cite{Maldacena:1997re}, where the correspondence between geometry and entanglement is a necessary instance of the gauge/gravity duality\cite{Swingle:2012wq,Bhattacharyya:2016hbx,Bao:2015uaa,Bao:2018pvs,Yang:2015uoa}, non-perturbative and background independent approaches to quantum gravity recently offered a different playground to investigate the roots of such interplay.

In loop quantum gravity~\cite{Rovelli:2008zza, thiemann_2007, rovelli_vidotto_2014}, and related generalizations~\cite{Perez:2012wv, Oriti:2013aqa}, like spin foam theory and group field theory, a beautiful quantization of space at one time, is realised in terms of \emph{quantum spin networks} ~\cite{Penrose_71, penrose_rindler_1984, Rovelli:1995ac}.  These are typically $SU(2)$ gauge symmetric tensor networks, defined by \emph{graphs} with edges labelled by $SU(2)$ spin \emph{irreps} and vertices dressed by intertwining operators \cite{rovelli_vidotto_2014}. 

For given graph, the spin-networks Hilbert space essentially capture the \emph{kinematics} of general relativity in its first order formulation~\cite{Ashtekar:1986yd}, which is eventually described in analogy with the conventional Hilbert space of an $SU(2)$ lattice Yang-Mills theory~\cite{Ashtekar:2014kba, Ashtekar:2021kfp}. 

In this sense, spin networks provide a phenomenal quantitative tool to investigate the operational content of the gravitational field, reduced on $3D$ spacetime slices, at the quantum scale~\cite{Bianchi:2012ev, Donnelly:2008vx, Donnelly:2011hn,Baytas:2018wjd,
Livine:2017fgq, Delcamp:2016eya, Bianchi:2016hmk,
Bianchi:2015fra, Chirco2018b, Chirco:2018fns, Oriti2018a}. Indeed, geometric and topological features of $3D$ geometry get encoded into purely algebraic and combinatorial degrees of freedom, hence local, non-local and holographic features of the entanglement structure of quantum spin networks are expected to reflect the signatures of Einstein's equations and the invariance for diffeomorphisms at the quantum level. 

In this light, spin networks entanglement has become a central resource to characterize physical vacuum states of the theory~\cite{PhysRevD.90.044044, Chirco_2015, Hamma:2015xla, Bianchi:2019pvv}, investigating \emph{local} holographic properties of quantum spacetime geometry \cite{Raju:2019qjq, Dittrich:2017hnl, Dittrich:2018xuk} and ultimately to study the emergence of classical spacetime geometry from its quantum description, with a renovate interplay of techniques and tools from quantum information theory, information geometry, quantum many-body theory and quantum computation theory more recently~\cite{ Livine:2006xk,Girelli:2005ii,Bianchi2016LoopEA, Feller:2017jqx,Anza:2016fix,Hohn:2017cpr, Chirco:2017xjb,Bianchi:2023avf,Sahlmann:2023plc, Haggard:2023tnj,Marchetti:2022nrf,Czelusta:2020ryq}. 

Most recent work in the field, in particular, has focused on quantum $3D$ geometry states with $2D$ boundaries, corresponding classically to space-time spatial slices with \emph{corners}~\cite{freidel2023corner}, thereby looking at quantum spin-networks with boundaries as \emph{boundary maps} with corner states encoding the geometric and topological information stored in the bulk correlation structure. In this setting, different measures of multipartite entanglement have been proposed to investigate encoding and decoding of bulk information in the boundary/bulk mapping~\cite{Chen_2021, Chen:2022rty, cepollaro2023} and its holographic behaviour. 

In~\cite{Chirco2018a,Chirco_2020,
Chirco:2021chk,Colafranceschi:2021acz,Colafranceschi:2022dig,Colafranceschi:2022ual},  the holographic character of the spin-networks boundary/bulk mapping has been investigated within a quantum typical regime starting from a \emph{random} tensor-network description of $3D$ quantum geometry states in the large spin limit ~\cite{Hayden:2016cfa,Qi:2013caa,Qi_2017}. Along this line, in \cite{cepollaro2023}, the authors first investigated the measure of \emph{entanglement negativity}, along the lines of \cite{Dong_21, Kudler_21, 38, Shapourian_21}, to extend the study of the hierarchy of boundary correlations to mixed states, beyond the bipartite pure-state setting of entanglement entropy.  

In this work, we consider a synthetic description of a uniformly curved quantum $3D$ space region with boundary, realised in terms of open \emph{tagged} spin-networks states in loop quantum gravity \cite{Livine:2013gna, Charles2016TheFS}. We show that the effective topology of this region is reflected in the multipartite entanglement of its \emph{tagged} spin-networks state description. In presence of intrinsic curvature, we generalise the bulk-to-boundary mapping description as to include the space of bulk defects in a generalised boundary space. In the resulting \emph{extended boundary} state, bulk information is \emph{shared} among entangled boundary spins and tags: surface and intrinsic curvature of the quantum space region are entangled. 

We model the generalised boundary mapping on a tripartite random state and compute the logarithm negativity for the bipartite reduced boundary, focusing on the universal typical behaviour of the large spin regime.

We find that the multipartite entanglement of the system depends not only on the bulk quantum correlations among the intertwiner states comprising the bulk of the spin network, but in an essential way, on the topology of the spin network state.

In particular, we show that the degree of quantum correlations among boundary subregions depends on the dimension of the bulk tags system, which plays the role of a hidden environment from the viewpoint of a generic observer measuring correlations on the boundary. When the bulk curvature environment is smaller than the boundary system, the entanglement negativity of the boundary subregions  displays the expected area law behaviour. However, when the environment is larger, the negativity vanishes suggesting an effective thermalization of the boundary. Remarkably, the PPT character of the mixed boundary state corresponds to change in the effective topology of the network in the large spin regime, with the two complementary boundary subregions becoming disconnected.

The paper is organized as follows. Section~\ref{sct:sn} introduces loop quantum gravity quantum geometry states as superpositions of quantum spin networks. In Section \ref{sct:sntop}, this notion is generalised to include boundaries and non-trivial topology. In particular, here we define quantum states associated with regions of bounded $3D$ space with non-vanishing intrinsic curvature. Section \ref{sct:tags} introduces the notion of tagged spin networks, which is the starting point of our characterisation of the bounded quantum geometry as a mixed state. 
In Section~\ref{sct:trip}, we define the tripartition $\{A_1, A_2, B\}$ of our system, with $A_1$, $A_2$ corresponding to complementary subregions of the boundary and $B$ referring to the set of tags produced by the integral (partial tracing) over the non-trivial loop holonomies in the bulk. For such a tripartite system, we quantify the entanglement of the boundary via a measure of \emph{typical} entanglement negativity~\cite{Peres_1996, PhysRevA.58.883,PhysRevA.60.3496, Eisert_99, PhysRevA.65.032314, Plenio_05, PhysRevB.94.035152} by considering the boundary mapping on a bulk state \emph{prepared at random}. The random measurement combined with the boundary map projection allows us to limit the partial tracing to the ``curvature'' degrees of freedom, while (partially) retaining the information on the graph structure. We compute typical $k$-th order R\'enyi log-negativity for even $k$ via replicas by mapping momenta of the $A_1, A_2$ reduced density matrix to partition functions of a classical generalised Ising model~\cite{Dong_21,Shapourian_21}, hence looking for the minimal free energy configurations of the model.
In Section \ref{sct:ising}, we characterise the bulk states contributions to the free energy of the classical model. The analysis is thereby restricted to states with small intertwiner entanglement.
Section \ref{sct:phases} contains an explicit example of open spin network state - a cluster of tags - with four vertices. For such a simplified setting, we can explicitly compute the typical logarithmic negativity of the mixed boundary state in the large spin regime and characterise the entanglement phase diagram of the boundary state. This section contains the main results of the paper. Finally, we close in Section \ref{sct:end} with a summary of the results and some closing remarks.

\section{Spin Networks states} \label{sct:sn}
Loop quantum gravity (LQG) space of 3D geometry is realized as a space of square-integrable functions, 
\begin{equation}
    H_\gamma = L^2\qty[SU(2)^{ \times E}/SU(2)^{\times V}]\, ,
\end{equation}
with support on closed oriented graphs $\gamma$ comprised by $V$ vertices connected by $E$ edges. States in $H_\gamma$ depend on one group element $g_e \in SU(2)$ for each edge $e$ of the graph and are assumed to be invariant under the $SU(2)$-action at each vertex $v$, that is, $\forall h \in SU(2)$,
\begin{eqnarray}
\psi_\gamma: SU(2)^{\times E}  &\to& \mathbb{C} \\ \nonumber
\{g_e\}_{e \in \gamma} &\mapsto& \psi_\gamma(\{g_e\}) = \psi_\gamma(\{h_{t(e)}\,g_e\, h_{s(e)^{-1}}\}) \,, \
\end{eqnarray}
where $t(e)$ and $s(e)$ respectively refer to the target and source vertices of the edge $e$ \cite{3}.

For $SU(2)$ is compact, by the Peter–Weyl theorem, functions of $d$ group elements $g$ can be decomposed into irreducible representations (irreps) of the group,
\begin{equation}
f(\{g\})=\sum_{\{j\}}\sum_{\{m\}\{n\}}  f^{\{j\}}_{\{m\}\{n\}}\prod_{i=1}^d d_{j_i} D^{j_i}_{m_i n_i}(g_i),
\end{equation}
with $\{j_i\}\in \mathbb{N}/2$ labelling irreps of $SU(2)$; indices $m_i$ ($n_i$) labelling a basis in the vector space $V^{j_i}=\text{span}\{\ket{j_i,m_i}\}$, carrying the  representation $j_i$, of dimension $d_{j_i}=2j_i+1$, and $D^{j_i}_{m_i n_i}(g_i)=\bra{j_i,n_i}g_i\ket{j_i,m_i}$ the Wigner matrix representing the group element $g_i$. 

This allows to write $SU(2)$-invariant functions $\psi_\gamma$ as \emph{superpositions} of gauge symmetric tensor networks  with fixed spins $\{j\}_{\gamma}$ and the connectivity of the graph $\gamma$. Two vertices of the graph, say $v$ and $w$, are connected by edge states labeled by a spin $j$ irreducible representation (irrep) of $SU(2)$ and dressed with a group element $g_e\in SU(2)$, that is
\begin{equation}
\ket{e_{vw} (g_e)} \equiv \sum_{\{m\}\{n\}} \frac{(-1)^{j-n}}{\sqrt{d_j}}\, D^{j}_{mn}(g_e)\, \ket{j,m}_v\otimes \overline{\ket{j,n}}_w 
\label{eq:edgehol}
\end{equation}
in the Hilbert space $H_{vw} = \Big[ V_{v}^{j}\otimes \overline{V}_{w}^{j}\Big]\,$ . At each vertex $v$ of the graph, an intertwiner operator $\iota_v$ enforces $SU(2)$ gauge invariance via a projection
\begin{equation}
\iota_v: \,\, \bigotimes_{e\in v} \, V_{j_e}\to V_0 \, .
\end{equation}
corresponding to a recoupling of the edges spins at the vertex in the $SU(2)$-symmetric (singlet) representation with $j=0$. This associates to each vertex the degeneracy space of $V_0$ for fixed $\{j_e\}$, 
\begin{equation}
H_v= \text{Inv}_{SU(2)}\Big[\bigotimes_{e\in v}
V^{j_e}\Big]= \text{span}\{\ket{\iota_v}\}\, ,
\end{equation}
with $\ket{\iota_v}$ an orthonormal basis in $H_v$. 

\begin{figure}[t!]
\includegraphics[width=0.40\textwidth]{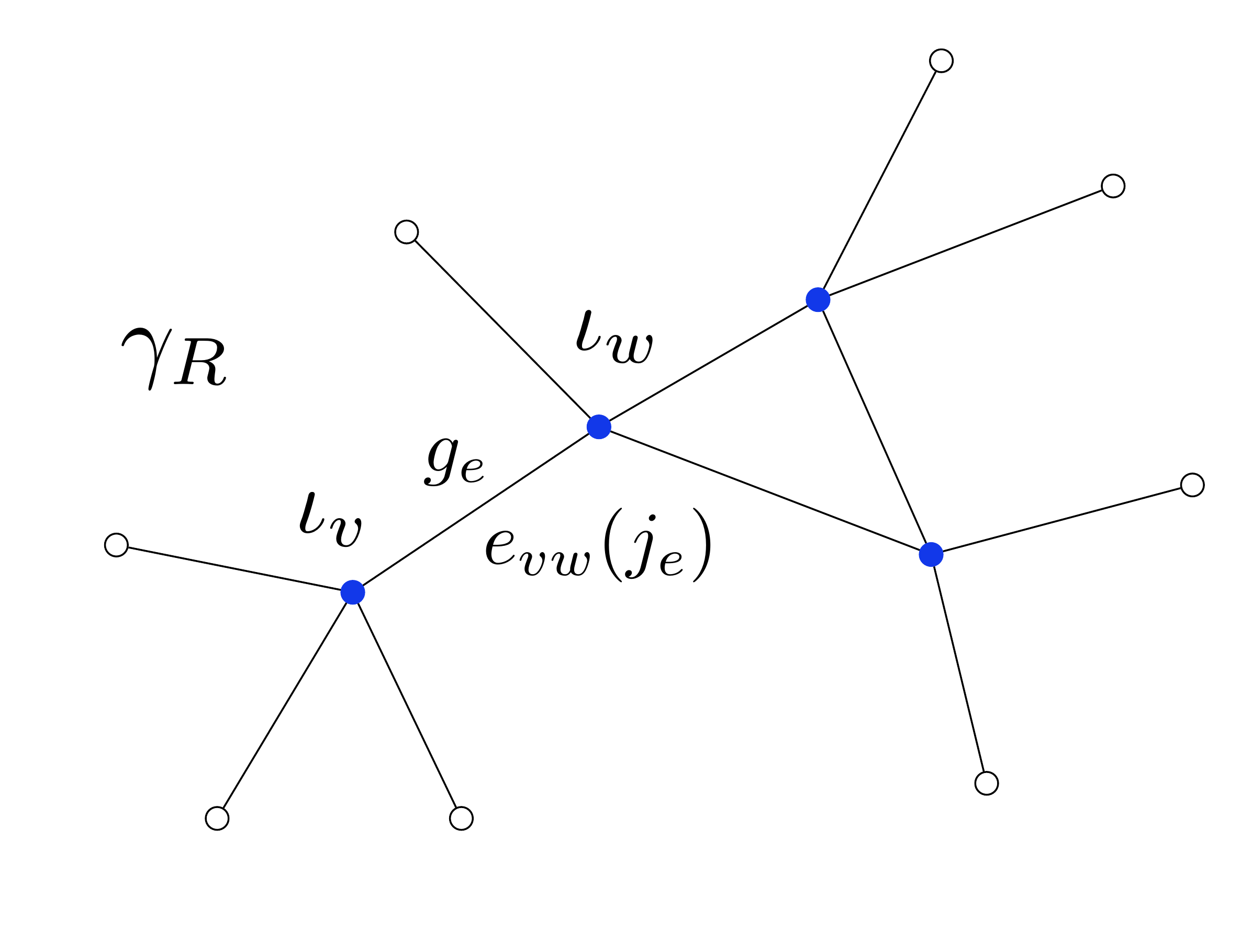}
\caption{Open spin network with support on the graph $\gamma_R$. The boundary is comprised by the set of uncontracted edges with dangling magnetic indices pictured as empty circles.}
\label{sn}
\end{figure}

A \emph{spin network} state $\ket{\gamma, \{j_e\},\{\iota_v\}}$ is thereby defined as the assignment of representation labels $\{j_e\}$ to each edge of $\gamma$ and the choice of a vector $\ket{\{\iota_v\}}$ in $H_V = \otimes_{v}^V H_v$ for the vertices, corresponding to the contraction (see Fig~\ref{sn})
\begin{equation}
\ket{\gamma, \{j_e\},\{i_v\}}=\left(\bigotimes_{e=1}^E \bra{e(g_e)}\right)\, \ket{\{\iota_v\}}.
\label{eq:projsn}
\end{equation}

Such spin-networks define an orthogonal basis in $H_\gamma$, 
for the following isomorphism holds
\begin{equation}
    H_\gamma\approx \bigoplus_{\{j\}} {H_V}.
\end{equation}
Any quantum $3D$ geometry state $\psi_\gamma$ can be constructed from the \emph{contraction} of a generic state $ \ket{\psi}$ in $H_V$ with the set of $E$ edge states, via a sum over spins
\begin{equation}
\ket{\psi_\gamma (\{g_e\})}= \bigoplus_{\{j_e\}}\left(\bigotimes_{e=1}^E \bra{e(g_e)}\right)\, \ket{\psi},
\label{eq:proj}
\end{equation}
where the contraction amounts to a sum over bulk indices.

\section{Spin-networks with boundary and Non-trivial Topology}\label{sct:sntop}
Imagine to cut a generic \emph{bounded} region $R$ out of quantum 3D space. The corresponding state will be defined with support on an \emph{open} graph $\gamma_R \subset \gamma$, which we can take to be comprised by 
the sets of vertices $\mathcal{V}_R$ and edges $\mathcal{E}_R$, with $\abs{\mathcal{V}_R}=V_R$ the number of vertices and $\abs{\mathcal{E}_R}=E_R$ the number of edges respectively. The \emph{boundary} of $R$, $\partial R$, is the set $\mathcal{E}_{\partial R}$ of $E_{\partial R}$ edges which have only one end attached to a vertex laying in $R$. Such boundary edges are characterised by an open (uncontracted) dangling index $m$, while bulk edges connecting vertices within $R$ are fully contracted. For states with support on open graphs, the contraction in \eqref{eq:proj} defines a \emph{mapping} of the state $\ket{\psi} \in H_{V_R}$ on the boundary space of uncontracted indices,
\begin{equation} 
{H}_{\partial R}=\bigoplus_{\{j_e\}}\bigotimes_{e \in \mathcal{E}_{\partial R}} V^{j_e}.
\end{equation}
In particular, the gauge invariant property of the LQG wavefunctions defined on closed graphs reduces to $SU(2)$-covariance of the bounded region wavefunction, 
\begin{equation}
\psi_R(\{g_e\}_{e \in \mathcal{E}}) = \psi_R(\{g_e\, h_{s(e)^{-1}}\}_{e \in 
\mathcal{E}_{\partial R}}) \,. \
\end{equation}
In the boundary mapping, the information on the bulk holonomies, as well as the correlation structure of $\psi\in H_{V_R}$ gets encoded in the coefficients of the resulting boundary state $\ket{\psi_R}$. 
We are interested in the information concerning the \emph{bulk curvature} and, in particular, in the way this affects (gets reflected in) the correlation structure of the boundary state $\ket{\psi_R}$. 

Bulk curvature is generally  associated to the presence of loops of non-trivial holonomies in the graph. However, the topology of the graph is only partially relevant to characterise such a degree of freedom. Indeed, thanks to the local gauge invariance of $\psi_R$, the structure of the bulk of $\gamma_R$ can be drastically simplified via a partial gauge-fixing of the bulk holonomies \cite{Livine:2006xk, Livine:2013gna, Charles2016TheFS}. In particular, for any open region $R$, the gauge-invariant Hilbert space on $\gamma_R$ is isomorphic to the gauge-invariant reduced space defined on a new graph $\Gamma_R$ consisting of a single vertex intertwining the external edges of the boundary $\partial R$ together with a number of (independent) loops $L$ fixed by the combinatorics of the region, that is $L=E_R-V_R +1$. If $L=0$ there are no loops and $\gamma_R$ has trivial topology. In particular, this implies that $R$ has vanishing \emph{intrinsic} curvature. 

More generally, the Hilbert space of the reduced graph $\Gamma_R$ will consist of the intertwiner space of the single bulk vertex times the product of the boundary representations spaces, that is 
\begin{multline}
\label{eq:projloopspace}
H_{\Gamma_R}=L^2[SU(2)^{\times (E+2L)}/SU(2)]\\
= \bigoplus_{\{j_{\ell,e}\}} H_v^{L} \otimes \bigotimes_{e\in \mathcal{E}_{\partial R}} V^{j_e} \, ,
\end{multline}
where the degeneracy space at the vertex now also involves the irreps on the loop 
\begin{equation}
H_v^{L} \equiv \text{Inv}_{SU(2)}\left[\bigotimes_{\ell=1}^{L} \left(V^{j_\ell} \otimes \overline{V}^{j_\ell} \right)\otimes \bigotimes_{e\in \mathcal{E}_{\partial R}} V^{j_e} \right]\,  . \label{eq:lint}
\end{equation}
The space in \eqref{eq:projloopspace}  provides a very synthetic description of the region $R$ in terms of a single ``loopy'' vertex with both its boundary degrees of freedom and bulk nontrivial holonomies. The gauge reduction isomorphism does not produce any coarse graining on physical degrees of freedom. However, the information of the combinatorics of the original graph is lost. As a consequence, the correspondence between the original graph and its flower is many-to-one \cite{Charles2016TheFS}.

\begin{figure}[t!]
\includegraphics[width=0.40\textwidth]{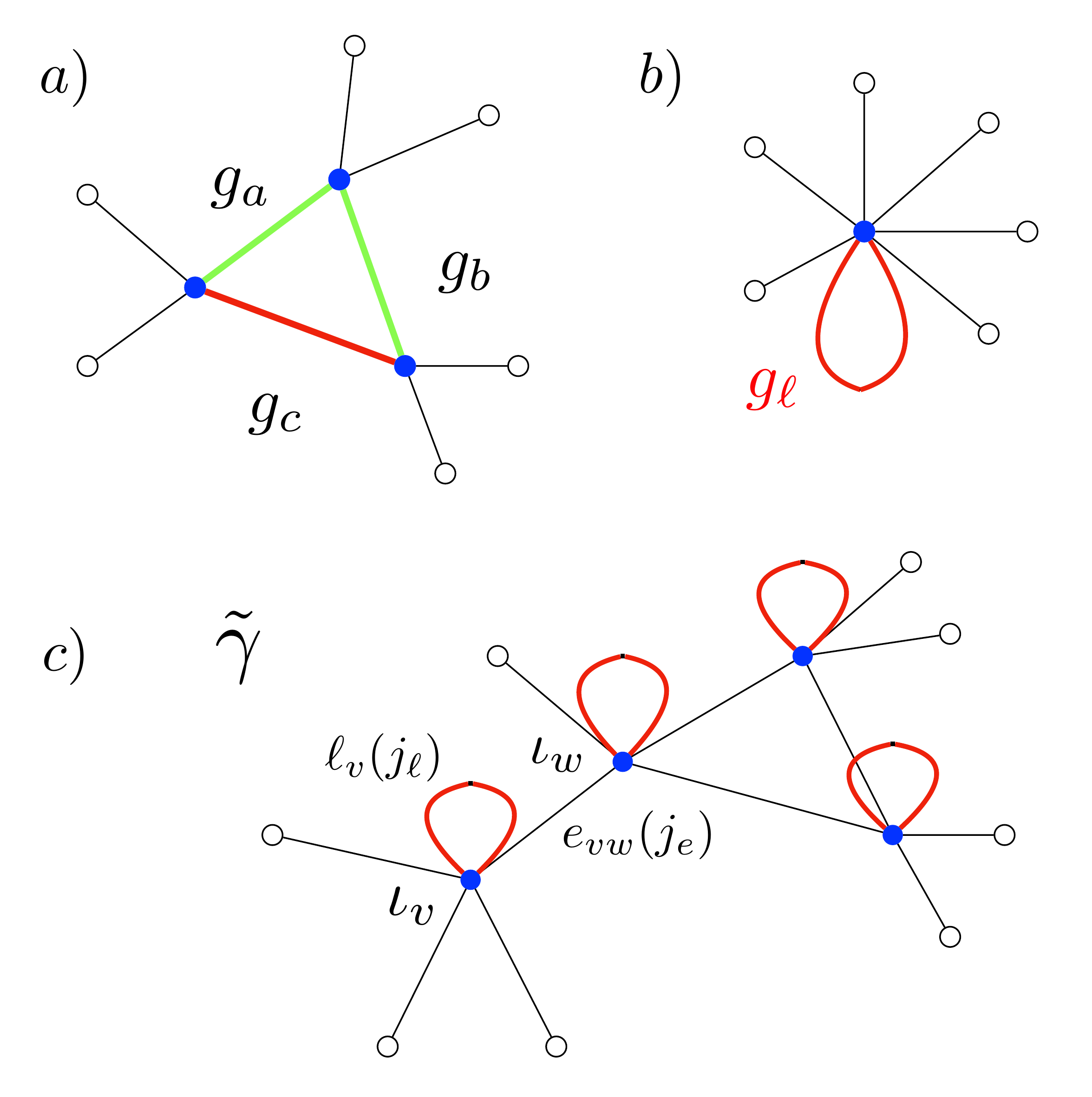}
\caption{a) Reduction by gauge fixing of the holonomies along the maximal spanning tree connecting three vertices of an open spin network; b) Resulting non-trivial holonomy on a loop attached to a coarse-grained vertex; c) Bounded quantum $3D$ region with uniformly distributed loops realised as the gluing of a set of coarse-grained vertices.}
\label{loop}
\end{figure}

Now, imagine to operate a similar reduction-by-gauge-fixing on a \emph{collection} of subregions, so as to end up with a set of loopy vertices glued together via edges dressed with a trivial holonomy. In this way, we construct a new spin network state with support on an open graph $\tilde{\gamma}_R$ associated to an extended $3D$ space region with  distributed \emph{intrinsic} curvature (see Fig.~\ref{loop}).
 Such an intermediate level of description allows to localize and keep track of the nontrivial holonomies $\{g_\ell\}$ (classically closure defects) responsible for the curvature at the vertices of the graph, without trivializing the topology of the graph, which is going to play a role in the entanglement structure of the boundary. 

\section{tags from spin-network kirigami}\label{sct:tags}

\begin{figure}[t!]
\includegraphics[width=0.40\textwidth]{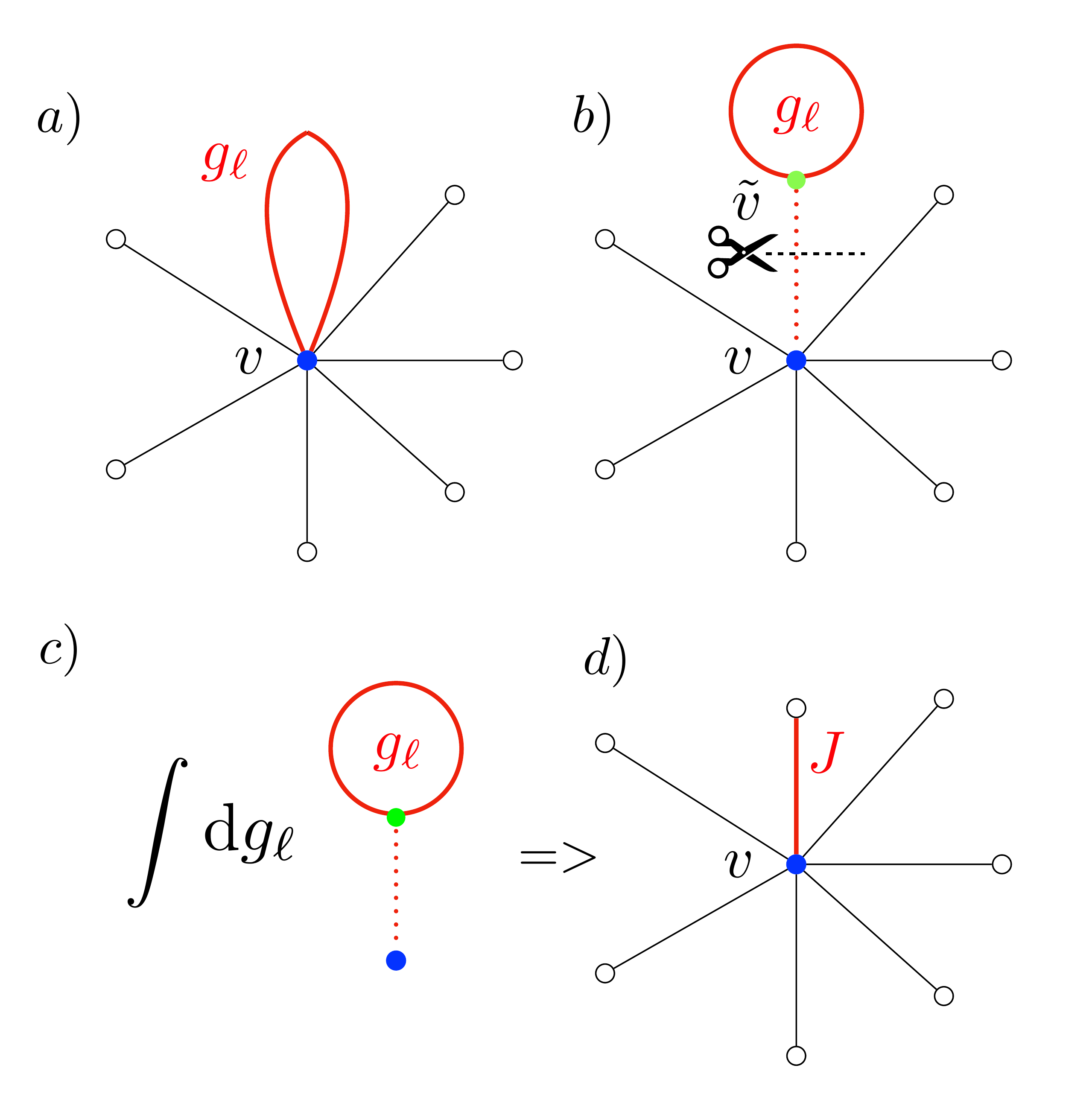}
\caption{a) A single loopy vertex ; b) vertex unfolding: isolate the nontrivial loop holonomy; c) partial trace via integration over the loop holonomy; d) tagged vertex.}
\label{kirigami}
\end{figure}

Along this line, let us consider for simplicity the case of a uniformly curved region of quantum $3D$ space, given by an open-graph state $\ket{\psi_R}=\ket{\psi_R(\{g_\ell\})}$ with all $V_R$ vertices carrying one single loop holonomy $g_\ell$. Further, let us assume for now that each vertex can have at most one open edge. This very simplified setting is what we need to neatly separate boundary from bulk degrees of freedom, as shown in Fig.\eqref{kirigami}. 
The first step in this sense consists in integrating out the loop holonomies. Following \cite{Livine:2019cvi}, for each spin $k$ carried by the loop, one can separate the loop from the edges at each vertex by unfolding the intertwiner into two different intertwiners connected by a virtual link labelled by an intermediate spin $J$: one intertwiner recoupling the boundary edges spins $j_e$ at $v$ into $J$ and a second 3-valent intertwiner recoupling the two copies of the spin k into $J$ at the virtual vertex $\tilde{v}$. Accordingly, the intertwiner space in \eqref{eq:lint} now decomposes as 
\begin{multline}
    H_v^{L} = \bigoplus_{J,\{k\}} \Bigg\{\text{Inv}_{SU(2)}\Big[\bigotimes_{e\in v}
    V^{j_e} \otimes V^J \Big] \\ 
      \otimes\, \text{Inv}_{SU(2)}\left[V^J \otimes \left(V^{k} \otimes \overline{V}^k \right) \right] \Bigg\}.
\end{multline}
A basis in $H_v^{L}$ is given by (see for instance~\cite{Livine:2017xww})
\begin{equation}
   \ket{J,k,\iota,\{j_e\}[g_\ell]} = D^k(g_\ell)\, C^{k,J|k}_{\tilde{m}, M|m}\,\braket{J,M}{\iota} 
   \label{eq:vtagbasis}
\end{equation}
with
\begin{equation}
C^{k,J|k}_{\tilde{m}, M|m}=\braket{k,m}{(k, \tilde{m}) \otimes (J,M)}
\end{equation}
the Clebsh-Gordan coefficient associated to the 3-valent intertwiner at $\tilde{v}$.

At each vertex of the graph, the integration over the loop holonomy only involves the Clebsh-Gordan coefficient and the Wigner matrix of the loop at the level of the basis in \eqref{eq:vtagbasis}. One can then compute the vertex density matrix $\rho_k$ as 
\begin{equation}
    \rho_k= \int \mathrm{d}g_\ell \ket{J,k,\iota,\{j_e\}[g_\ell]} \bra{J,k,\iota,\{j_e\}[g_\ell]}
\end{equation}
and show that the sum of $\rho_k$ over the loop spin $k$ gives exactly the identity matrix $\rho= \braket{J,M}{\iota} \braket{\iota}{J,M}$, with $\ket{\tau}=\braket{J,M}{\iota}$ a \emph{basis} in the \emph{reduced} vertex space ~\cite{Livine:2017xww}
\begin{equation}
     H_v^J = V^J \otimes  M_J^{\{j_e\}} \, ,
\end{equation}
with $M_J^{\{j_e\}} \equiv \text{Inv}_{SU(2)}\Big[V^J \otimes \bigotimes_{e \in v} V^{j_e}\Big]\,$. 

As a result of the integration, curvature is locally encoded into a virtual spin $J$, or \emph{tag}, attached to the vertex. We can see the vector $\ket{J, M} \in V^J$ as the result of the recoupling of all the spins $j_e$ at the vertex into a non-vanishing overall spin $J$, a topological defect which breaks local gauge invariance.\footnote{Given a tagged basis $\ket{\tau_v}$ at each vertex, we can define a \emph{tagged} \emph{spin network} basis 
 \begin{equation}
 \ket{\gamma, \{j_e\},\{\iota_v\}, \{J_v\}}=\left(\bigotimes_{e=1}^{E} \bra{e(g_e)}\right)\, \ket{\{\tau_v\}}
\label{eq:projtag}
 \end{equation}
 for the whole state on $\tilde{\gamma}_R(V_R, E_R, E_{\partial R})$.}

Starting from the tagged intertwiner description, we now define a \emph{homogeneously curved} quantum $3D$ bounded space geometry as the \emph{extended} boundary map
 \begin{equation}
 \ket{\psi_\tau}=\left(\bigotimes_{e=1}^{E} \bra{e(g_e)}\right)\, \ket{\psi_b}
\label{eq:projt}
 \end{equation}
 where the generic tagged state $\ket{\psi_b}$ is defined, for fixed spins $\{J, j_e\}_v$, as
\begin{equation}\label{eq:tag-state}
\ket{\psi_b}=\sum_{ \{M_v\} \{\iota_v\}} C(\{J,j_e\})_{\{M_v\},\{\iota_v\}} \bigotimes_{v \in \mathcal{V}_R}\ket{\tau_v}
\end{equation}
in the tensor product space of $V_R$ tagged intertwiner spaces $H_{V_R} \equiv \bigotimes_{v}^{V_R} H_v^{J_v}$. Differently from \eqref{eq:proj}, the projected state now lives in an \emph{extended} boundary space, 
\begin{equation}\label{eq:projtagspace}
H_\tau\equiv
\bigotimes_{v\in \mathcal{V}_R} V^{J_v} \otimes \bigotimes_{e\in \mathcal{E}_{\partial R}} V^{j_e}  \, 
\end{equation}
which comprises the tensor product of tags and boundary-spins representation spaces.\footnote{In fact, one can think of the tag factor in \eqref{eq:projtagspace} as a disjoint \emph{inner} boundary associated to closure defects at the bulk vertex. This is compatible with a picture of the tags as the result of a set of holes cut out of planar spin network punctured by uncontracted bulk edges. In the standard picture, such holes correspond to loops which recouple the puncturing edges.} 

The tagged open graph state is written as
\begin{multline}\label{eq:projtag-state}
\ket{\psi_\tau}=\sum_{\{M_v, m_e\}} \mathcal{C}(\{\iota_v, J_v, j_e\})_{\{M_v, m_e\}}\\ \bigotimes_{v \in \mathcal{V}_R}\ket{ J_v,M_v} \bigotimes_{e \in \mathcal{E}_{\partial R}}\ket{j_e,m_e},
\end{multline}
with coefficients $ \mathcal{C}(\{\iota_v, J_v, j_e\})_{\{M_v, m_e\}}$ encoding the information on the quantum correlations among bulk intertwiners, the connectivity, and the topological defects of the graph.
We expect states in \eqref{eq:projtag-state} to be generically highly entangled, with information about bulk geometry encoded into the boundary state coefficients $\mathcal{C}$ in a complicated way. In the extended boundary system, such information is shared by boundary spins and bulk tags. 

Now, consider an observer having access only to the actual boundary system. Such an observer is precluded from measuring the information on the bulk, hence they will typically describe the system via a mixed boundary state. Concretely, this amounts to a partial tracing over the bulk. In our analysis, we describe the coarse-grained viewpoint of the boundary observer by introducing a \emph{random measurement} on the state $\ket{\psi_\tau}$. We then investigate the boundary-system/bulk-environment coupling via a combination of random projections and partial tracing over the tags.

\section{Negativity of a tripartite boundary-tags system} \label{sct:trip}

Let us consider the extended boundary space \eqref{eq:projtagspace}. For convenience, let us assume curvature, in the form of closure defects, to be homogeneously distributed throughout the graph $\tilde{\gamma}_R$. Accordingly, we set all the spins of the graph $\{j_e\}$ and $\{J_v\}$ to be equal to the single value $j$ and $J$ respectively, generally with $j\neq J$. 

A \emph{random measurement} on the bulk state $\ket{\psi_b}$, prepared on $\bigotimes_v\, H_v^J$, is realized by a projection onto a set of independent and individually (Haar) random vertex states $\ket{f_v}$ in the \emph{extended} vertex space
\begin{equation}\label{eq:extvert}
K_v^J=V^J \otimes \bigotimes_{j_e \in v} V^{j_e }\otimes  M_J^{\{j_e\}} \, ,
\end{equation}
which comprises the intertwiner space at the vertex, the spin irrep spaces of the edges connecting the vertex with the rest of the graph, as well as the tag spin space
. In particular, we can obtain the state $\ket{f_v}$ by acting on a reference state $\ket{0}$ in $K_v^J$ at $v$ with a Haar random unitary, that is $\ket{f_v}= U_v \ket{0}$ \cite{Hayden:2016cfa}. 

The generic random vertex state is explicitly written as 
\begin{equation}
\ket{f_v}= \sum_{\iota_v,\{n_e\}} f^{v}_{\iota_v,\{n_e\}} \, \braket{J,M}{\iota_v} \otimes \bigotimes_{e\in v}\ket{j_e,n_e}\, 
\label{rvstate}
\end{equation}
The randomised extended boundary density matrix is then simply defined along with \eqref{eq:projt} by a trace over bulk indices, which contract $\rho_b= \ketbra{\psi_b}$ and $\rho_E= \ketbra{E}$, for $\ket{E}= \bigotimes_e^{E} \ket{e}$, with the further insertion of the \emph{random projector} $\Pi= \bigotimes_v^{V_R} \Pi_v=  \bigotimes_v^{V_R} \ketbra{f_v}$: 
\begin{equation}\label{eq:rndmap}
{\rho}_\tau = \text{Tr} \left[\,\rho_b \,\cdot \, \rho_E  \, \Pi \right]\, .
\end{equation} 
The bulk state $\rho_b$ and the maximally entangled edges state $\rho_E$ enter as product states in the trace, while the information on the tags is enclosed in the bulk state. 

We shall define a tripartition of the extended boundary system into three subsystems $A_1$, $A_2$, $B$, with $A_1$, $A_2$ complementary regions of the external boundary, and $B$ the set of tags in the bulk. This corresponds to the following factorization of the extended boundary Hilbert space {\begin{equation}\label{eq:tripa}
H_\tau =  \bigotimes_{e\in A_1}V^{j_e} \otimes \bigotimes_{e\in A_2}V^{j_e}\otimes \bigotimes_{i\in B} V^{J_i}\,.
\end{equation}}
Starting from \eqref{eq:tripa}, we focus on the reduced state obtained tracing over the tag spins in $B$,
\begin{equation}
\rho_{A_1 A_2}= \text{Tr}_B [\rho_\tau]\,.
\label{eq:roab}
\end{equation}
We evaluate the entanglement of the mixed state $\rho_{A_1 A_2}$ by computing the \emph{logarithmic negativity} of the \emph{partial transpose} of the state \cite{Donnelly:2008vx, 20, Dong_21, Shapourian_21, cepollaro2023}. In the bipartite subsystem $A_1A_2$, with Hilbert space ${H}_{A_1}\otimes {H}_{A_2}$, given an orthonormal basis $\ket{i}_{A_1}$ in $\mathcal{H}_{A_1}$ and similarly $\ket{j}_{A_2}$ in $\mathcal{H}_{A_2}$, the { partial transpose} of $\rho_{A_1A_2}$ with respect to the subsystem $A_2$ is defined as 
\begin{equation}
(\rho_{i_{A_1}j_{A_2},k_{A_1}l_{A_2}})^{T_{A_2}}=\rho_{i_{A_1}l_{A_2},k_{A_1}j_{A_2}}\,.
\end{equation}
The eigenvalues of the partially transposed reduced density matrix $\rho_{A_1A_2}^{T_{A_2}}$ are real since the partial transposition is an Hermitian and trace-preserving map. Yet $\rho_{A_1A_2}^{T_{A_2}}$ is not completely positive, \emph{i.e.} it may have negative eigenvalues. The presence of negative eigenvalues in the partial transpose \emph{witnesses} quantum correlations in $\rho_{A_1A_2}$. 
Concretely, the amount of entanglement of the state can be quantified by counting the number of negative eigenvalues of $\rho_{A_1A_2}^{T_{A_2}}$ \cite{Plenio_05}, as
\begin{equation}
N(\rho_{AB})\equiv\frac{\lVert\rho_{AB}^{T_B}\rVert_1 -1}{2}=\sum_{i:\lambda_i<0}\abs{\lambda_i} \, ,
\end{equation}
where $\lVert \cdot\rVert_1$ is the trace norm, and the logarithmic negativity is defined as
\begin{equation}\label{neg}
    {E}_N(\rho_{A_1 A_2})\equiv \log \lVert\rho_{A_1A_2}^{T_{A_2}}\rVert_1 \, .
\end{equation}
Both $N$ and $E_N$ are entanglement monotone under general positive partial transpose (PPT) preserving operations \cite{Peres_1996}. 
In order to compute \eqref{neg}, the first key ingredient is to first look at the $k$-th R\'enyi negativity of $\rho_{A_1 A_2}$,\footnote{Notice that the trace operation $\text{Tr}$ in \eqref{negz} now involves sums over boundary indices (dangling edges and tags). The denominator in \eqref{negz} directly comes from the normalization of $\rho_{A A_2}$.} 
\begin{equation}
N_k(\rho_{A_1 A_2})=\Tr \qty[ \qty(\rho_{A_1 A_2}^{T_{A_2}})^k\Big/ \qty(\text{Tr}\qty[\rho_{A_1 A_2}])^k]\, , \label{negz}
\end{equation}
hence, eventually recover the logarithmic negativity by taking the $k\to 1$ limit of the logarithm of the analytic continuation of the momenta for even $k$ \cite{Dong_21}.
In particular, since we took a random measurement on the bulk state, we will need to compute the R\`enyi negativity in \emph{expected value}, with respect to the uniform Haar measure $\mu$, namely
\begin{equation} 
\mathbb{E}_{\mu} \left[ N_k(\rho_{A_1 A_2})\right] \equiv \overline{N_k(\rho_{A_1 A_2})}\, .
\end{equation} 
Following \cite{Dong_21, Hayden:2016cfa, cepollaro2023}, we know that, due to concentration of the trace, a regime of large spins allows us to approximate the R\`enyi negativity by the ratio of expected values of the $k$-th moment and the $k$-th power of the partition function of ${\rho}_{A_1 A_2}^{\,T_ {A_2}}$, that is
\begin{equation}
\overline{N_k(\rho_{A_1 A_2})} \simeq \frac{\overline{\Tr\left[\qty({\rho}_{A_1A_2}^{\,T_{A_2}})^k\right]}}{\overline{\left(\Tr\left[\qty({\rho}_{A_1A_2})\right]\right)^k}} \equiv \frac{\overline{Z^{(k)}_1}}{\overline{Z^{(k)}_0}}\,.
\label{454}
\end{equation} 
Moreover, in the large spin regime, we have 
\begin{equation}
N_k(\rho_{A_1 A_2}) \simeq \overline{N_k(\rho_{A_1 A_2})}\, ,  
\end{equation} 
that is, the function is well approximated by its \emph{typical value}.

\begin{figure}[t!]
\includegraphics[width=0.4\textwidth]{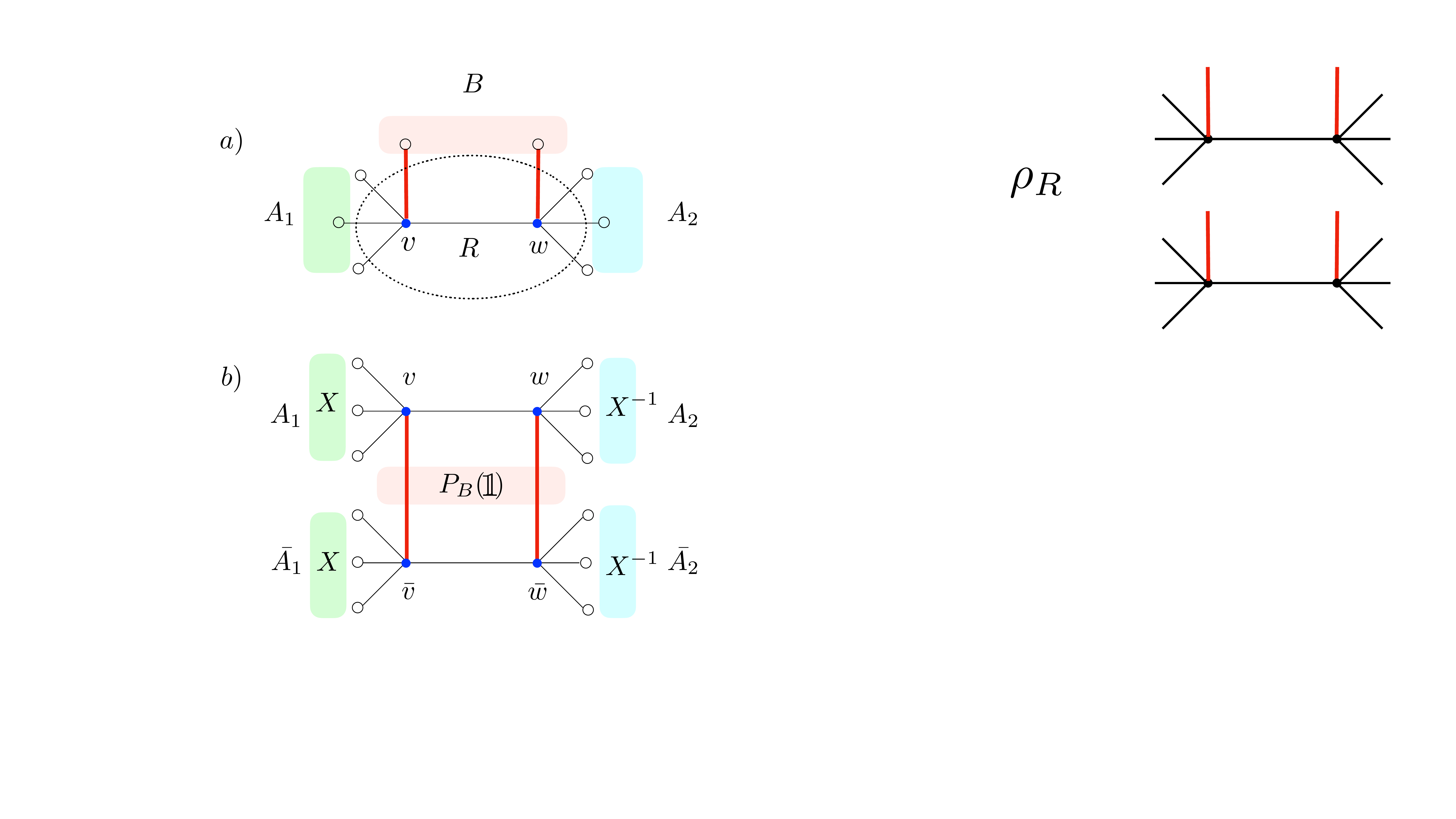}
\caption{a) Tripartition of the spin network extended boundary for a simple graph comprised by two 4-valent vertices glued via an edge; b) Partial tracing of the density matrix over the  tags space. }
\label{trip}
\end{figure}

The expected values of the partition functions in \eqref{454} are computed via the standard \emph{replica} technique. We first linearize the partial transpose matrix as follows,
\begin{multline}
\Tr \left[ \left({\rho}_{A_1A_2}^{\,T_{A_2}}\right)^k\right]= \Tr \left[{\rho}_{A_1A_2}^{\,\otimes k} \, P_{A_1}\qty(X)\otimes P_{A_2} \qty(X^{-1})\right]\\
\qquad = \text{Tr}\left[{\rho}_{R}^{\,\otimes k}\, P_{A_1}\qty(X) \otimes P_{A_2}\qty(X^{-1})\otimes P_B\qty(\mathbbm{1})\right]\, .
\end{multline}
Here, we generally denote $P_{I}(\sigma)$ as a unitary representation of the permutation $\sigma$, {$I=A_1,A_2,B$}, with $X$, $X^{-1}$ and $\mathbbm{1}$ the cyclic, anti-cyclic and identity permutations. On the spins and tags dangling indices, the operator $P_B\qty(\mathbbm{1})$ realises the partial trace over the tags on $\rho_b$, while $P_{A_1}\qty(X) \otimes P_{A_2}\qty(X^{-1})$ act on the $k$ copies of uncontracted spins in $\rho_E^{\otimes k}$ implementing the replica trick and the partial transposition respectively. 

For the linearity of the trace, the expectation value can be carried out before taking the trace and it will concern only the random tensors. From \eqref{eq:rndmap}, we have
\begin{equation}\begin{split}
\overline{Z_1^{(k)}}&=\Tr\Bigg[\rho_b^{\otimes k}\otimes\rho_E^{\otimes k}\, \qty(\bigotimes_v\overline{\qty(\ket{f_v}\bra{f_v})^{\otimes k}})  \\
\quad & \cdot P_{A_1}\qty(X)\otimes P_{A_2}\qty(X^{-1})\otimes P_B\qty(\mathbbm{1})\Bigg]\, , \\
\overline{Z^{(k)}_0}&=\Tr\qty[\rho_b^{\otimes k}\otimes\rho_E^{\otimes k}\, \qty(\bigotimes_v\overline{\qty(\ket{f_v}\bra{f_v})^{\otimes k}})]\, ,
\end{split}\label{parti}
\end{equation}
where the overall trace now runs over both bulk and boundary indices.

By Schur's lemma, the average of the $k$ copies of the vertex state in \eqref{parti} results in a sum over unitary representations of the permutation group $g_v$ acting on the $k$ copies of the extended vertex space $K_v^J$ \cite{38, Dong_21, cepollaro2023},
\begin{equation}
\overline{\qty(\ket{f_v}\bra{f_v})^{\otimes k}}=\frac{\qty(D_v-1)!}{\qty(D_v+k-1)!}\sum_{g_v\in S_k}P_v(g_v)\, ,
\end{equation}
with dimension $D_v= \text{dim}(K_v^J)$.  

By performing the average, individually on each independent random vertex, we obtain\footnote{ $\overline{Z^{(k)}_0}$ has the same form with $X$ and $X^{-1}$ replaced by $\mathbbm{1}$.}
\begin{multline}\label{z1 with permutations}
\overline{Z^{(k)}_1}=\mathcal{C}\Tr\Bigg[\rho_b^{\otimes k}\otimes\rho_E^{\otimes k}\, \qty(\bigotimes_v \sum_{g_v\in S_k}P_v(g_v))  \\ \quad \cdot  P_{A_1}\qty(X) \otimes P_{A_2}\qty(X^{-1})\otimes P_B\qty(\mathbbm{1})\Bigg]\, , \end{multline}
with $\mathcal{C} = \prod_v\qty[\frac{\qty(D_v-1)!}{\qty(D_v+k-1)!}]$.

\begin{figure}[t!]
\includegraphics[width=0.4\textwidth]{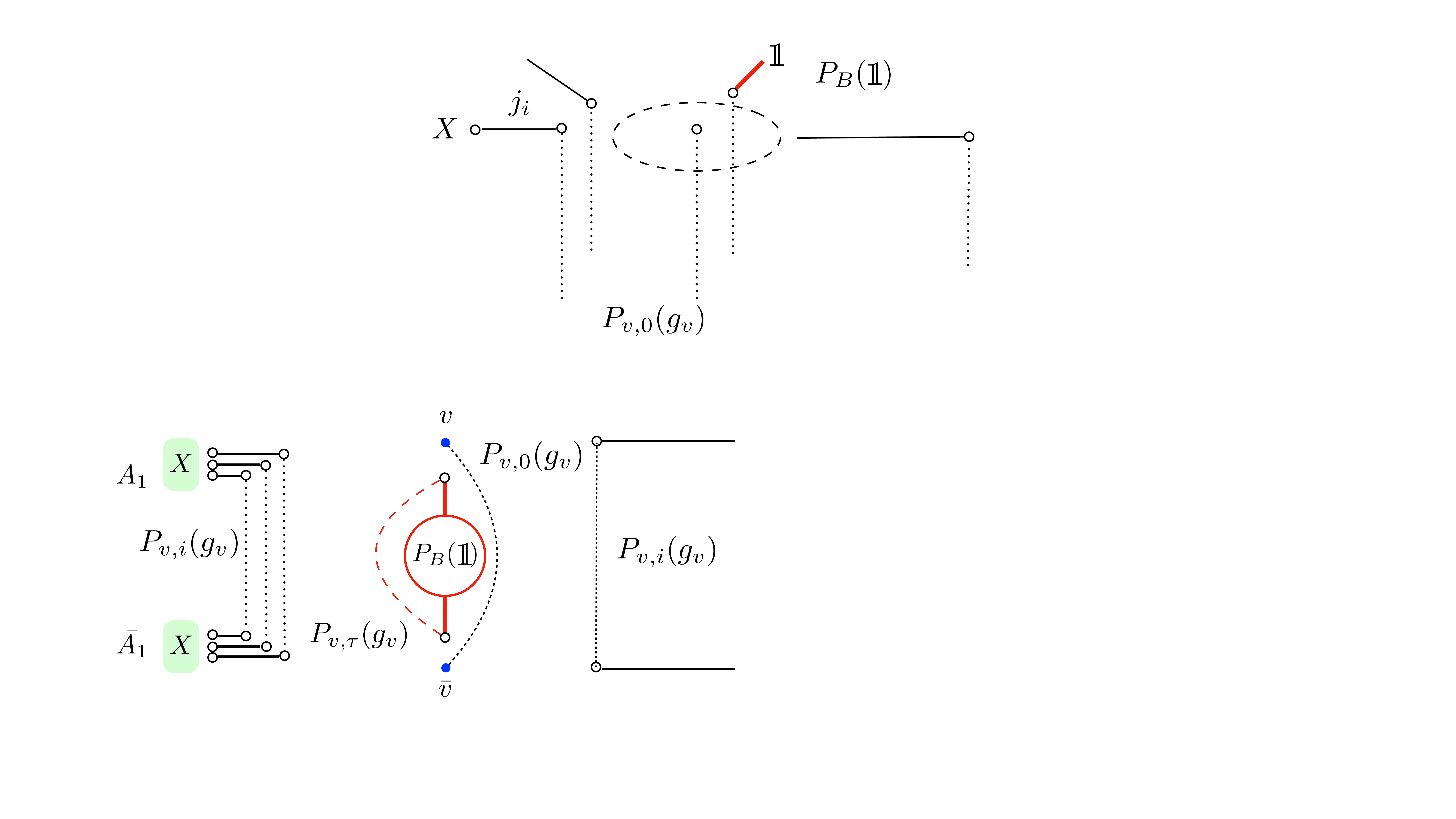}
\caption{Permutation operator $P_v(g_v)$ action factorizes on the indices of the extended vertex state. In the example in figure the operator acts as the identity connecting the set of internal indices of $v$ to the same set in $\bar{v}$, while $P_B\qty(\mathbbm{1})$ realises the partial trace over the tags.}
\label{pro-v}
\end{figure}

On the bulk indices at each vertex, the permutation operator $P_v(g_v)$ factorizes into \emph{four} independent operators (see fig.~\ref{pro-v}),
\begin{multline}
    P_v(g_v)= P_{v,0}(g_v)\, \otimes \bigotimes_{e^i_{vw} \in \mathcal{E}_R} P_{v,i}(g_v) \\ 
    \otimes \bigotimes_{e^i_{v\bar{v}} \in \mathcal{E}_{\partial R}} P_{v,i}(g_v)\, \otimes P_{v,\tau}(g_v)\, ,\label{permvert}
\end{multline}
where $P_{v,0}(g_v)$ acts on $k$ copies of the multiplicity intertwiner space, {$ \bigotimes_{e^i_{vw} \in \mathcal{E}_R}P_{v,i}(g_v)$} acts on $k$ copies of the internal edges, $\bigotimes_{e^i_{v \bar{v}} \in \mathcal{E}_{\partial R}} P_{v,i}(g_v)$ acts on the boundary semi-edges,\footnote{ We indicate by $\bar{v}$ a virtual vertex connected to $v$ by the open boundary edge $e^i_{v\bar{v}}$.} and finally $P_{v,\tau}(g_v)$ acting on $k$ copies of the \emph{recoupled tag spin at the vertex}.

Accordingly, the trace in \eqref{z1 with permutations} factorizes over the Hilbert spaces of a) internal edges, b) boundary spins, c) tags spins and d) bulk intertwiners. As a result, we can eventually rewrite the $k$th power of the normalized partition function in \eqref{454} as 
\begin{equation} 
\frac{\overline{Z^{(k)}_1}}{\overline{Z^{(k)}_0}}= \sum_{\{g_v\}} e^{-A^{(k)}_{1/0}}\, , \label{Partition}
\end{equation}
where $A^{(k)}_{1/0}\equiv A^{(k)}_1/A^{(k)}_0$ defines the (normalized)\footnote{with 
   \begin{multline}\label{A0 bulk generico}
   A^{(k)}_0 = \sum_{e^i_{vw} \in \mathcal{E}_R}\Delta(g_v,g_w)\log d_{j^i_{vw}} \\   + \sum_{e^i_{v\bar{v}} \in A_1 \cup A_2} \Delta(g_v, \mathbbm{1})\log d_{j^i_{v\bar{v}}} +  A_b + \xi\, .
\end{multline}} action of a classical \emph{generalized} Ising-like model:

\begin{multline}\label{A1 bulk generico}
    A^{(k)}_1 \big[\{g_v\}\big] = A_{b} + \xi +\sum_{e^i_{vw} \in \mathcal{E}_R}\Delta(g_v,g_w)\log d_{j^i_{vw}}  \\
    +\sum_{e^i_{v\bar{v}} \in A_1} \Delta(g_v, X)\log d_{j^i_{v\bar{v}}} +  \sum_{e^i_{v\bar{v}} \in A_2} \Delta(g_v, X^{-1})\log d_{j^i_v}\,,
   \end{multline}
   with $\xi$ a constant term and \begin{multline}
    A_{b} = -\log \Bigg \{ \Tr \Bigg[\rho_b^{\otimes k}\otimes P_B\qty(\mathbbm{1})\, \cdot\\ 
     \cdot\qty(\bigotimes_v P_{v,0}(g_v)^{\,} \otimes P_{v,\tau}(g_v))\Bigg]\Bigg\}\label{eq:bterm}
\end{multline}
the bulk state contribution to the action, which accounts for the presence of curvature measured by the tags term in $\rho_b$. 
In the analogy, the \emph{generalized spins} correspond to the permutation operators $ P_v(g_v)$ sitting on the vertices of the open spin network graph $\tilde{\gamma}_R$ acting on the $k$ replicas of the vertex indices space. The permutations fields $X$, $X^{-1}$ and $\mathbbm{1}$ play the role of boundary conditions set on the virtual one-valent vertices ($\bar{v}$) attached to the dangling spins and tags of the extended boundary.

The two-body interaction terms measure the length of the permutation-loops across the replicas, weighted by the (log of the) dimension of the representation carried by the edges of the graph. These lengths are conveniently described as Cayley distances $\Delta\qty(g,h) = k - \chi(g^{-1}\,h)$ on the permutation group $S_k$, where $\chi(g)$ indicates the number of cycles in a permutation $g$ \cite{Dong_21, cepollaro2023}.\footnote{This is $k$ for the identity and $1$ for a $k$-cycle. Then, we see that $\Delta\qty(g,h)$ gives the minimal number of swaps to go from $g$ to $h$, for $g,h \in S_k$.}  

The calculation of typical value of Rényi negativity is then mapped to the minimization of the action \eqref{A1 bulk generico}. In general, the three boundary conditions will propagate inside the spin network graph $\tilde{\gamma}_R$. Depending on the couplings and the connectivity of the graph, different configurations of the model will be associated to different sets of domains for $X, X^{-1},\mathbb{1}$ in the bulk. As usual, we expect maximal domains to be associated to equilibrium configurations. In particular, maximal domains correspond to minimal domain walls.

By describing interactions via a Cayley distance we see that the Rényi negativity should depend directly on the \emph{area} of such minimal domain walls in the network, given by the number of edges of $\tilde{\gamma}_R$ associated to a non-vanishing Cayley distance between permutations in the two domains. 
In particular, thanks to the existence of a well defined loop quantum gravity area operator acting on the edges of the spin network, we can map domain walls areas to actual quantum geometry surfaces in our $3D$ space region $R$.

\section{Generalized Ising Model: modelling Bulk contributions}\label{sct:ising}

The tensor product form of the edges and bulk density matrices in \eqref{z1 with permutations} corresponds to a factorization of the typical R\'enyi negativity into a combinatorial contribution associated to the distribution of maximally entangled edge states $e_{vw}$ and a bulk contribution carrying physical correlations among intertwiners. The latter, in particular, contains the information on the intrinsic curvature of the $3D$ geometry, in the form of tags, as apparent from \eqref{eq:tag-state}. We would expect such information to be highly mixed in the bulk. However, after the partial tracing over the loop holonomies, the $SU(2)$-covariance of the tagged-bulk state induces $SU(2)$-invariance on the associated density matrix, which necessarily reads as a tensor product for fixed tags spins $\{J_v\}$, that is 
\begin{equation}
\rho_b=\bigotimes_v^{V_R} \frac{\mathbbm{1}}{2J_v+1} \otimes \rho_{\{J_v\}} \label{ete}
\end{equation}
which corresponds to a totally mixed state on the tags space and a generally non-trivial density matrix $\rho_{\{J_v\}} \in \text{End}[H_{V_R}]$~\cite{Livine:2017xww,Chen:2021vrc}. 

From the form of $\rho_b$ in \eqref{ete}, we have
\begin{equation}
\rho_b^{\otimes k}=\bigotimes_{v=1}^{V_R} \qty(\frac{\mathbbm{1}}{2J_v+1})^{\otimes k} \otimes \rho_{\{J_v\}}^{\otimes k}  \, ,
\end{equation}
then the bulk term $A_b$ generally factorizes as follows
\begin{align} \label{eq:bterm1}
A_{b} &= -\log \Bigg \{ \Tr \left[\rho_{\{J_v\}}^{\otimes k} \otimes \bigotimes_v  \qty(\frac{\mathbbm{1}}{2J_v+1})^{\otimes k} \right. \\ \nonumber  
    &\qquad  \left . \otimes \, P_B\qty(\mathbbm{1}) 
  \left(\bigotimes_v P_{v,0}(g_v)^{\,} \otimes P_{v,\tau}(g_v)\right)\right]\Bigg \} \\ \nonumber  
    &\qquad = A_\iota +A_\tau  
\end{align} 
with a bulk intertwiner contribution $A_\iota$ defined as
\begin{equation}
 \begin{split}
A_\iota\equiv -\log \left\{ \Tr \qty[ 
\rho_{\{J_v\}}^{\otimes k}  \qty(\otimes_v P_{v,0}(g_v))] \right\} \, ,
\end{split}
\end{equation}
and a contribution of the tags, $A_\tau$, given by
\begin{multline}
A_\tau \equiv -\log \Bigg \{ \Tr \Bigg[ \bigotimes_v  \qty(\frac{\mathbbm{1}}{D_{J_v}})^{\otimes k}  \otimes \, P_B \qty(\mathbbm{1}) \, \cdot \\   
\cdot \, \big(\otimes_v P_{v,\tau}(g_v)\big)\Bigg] \Bigg \} \, ,
\label{eq:tgscont}
\end{multline}
where we used $D_{J_v}=2J_v+1$, the dimension of the tag space. The result of the trace in \eqref{eq:tgscont} is
 \begin{multline}
\qty(\prod_{v\in \mathcal{V}_R} D_{J_v}^{-k})\Tr \qty[\bigotimes_{\langle v \bar{v}\rangle} \qty(\sum_{g_v\in S_k}P_v(g_v))\otimes P_B(\mathbbm{1})]\\
    =\sum_{\{g_v\}} \prod_v D_{J_v}^{-\Delta(g_v,\mathbbm{1})}\, .
\end{multline}
Therefore, the tags contribution to the action reads
\begin{equation}
A_\tau = \sum_{v} \Delta(g_v, \mathbbm{1})\log D_{J_v} \, ,
 \end{equation}
 which ends up being equivalent to a pairwise interaction term from an inner edge. 
The full action can be written as
\begin{eqnarray}
    &&A^{(k)}_1 =\xi +  A_\iota + \sum_{e^i_{vw} \in \mathcal{E}_R}\Delta(g_v,g_w)\log d_{j^i_{vw}} \nonumber \\
    &&+\sum_{e^i_{v\bar{v}} \in A_1} \Delta(g_v, X)\log d_{j^i_v} +  \sum_{e^i_{v\bar{v}} \in A_2} \Delta(g_v, X^{-1})\log d_{j^i_v} \nonumber\\ 
    &&+ \sum_{v} \Delta(g_v, \mathbbm{1})\log D_{J_v} \label{A1-full}
   \end{eqnarray}
The bulk contribution $A_\iota$ is nontrivial as long as the bulk intertwiners are correlated. In fact, it vanishes only if we consider $\rho_{J_v}$ to be a product state in $\text{End}[H_{V_R}]$ as shown in \cite{cepollaro2023}. On the other hand, both edges and tags contributions have to do with the topology of the graphs support $\tilde{\gamma}_R$, connectivity and topological defects respectively.

As a general result of our analysis, we can then formally divide the action in three terms as follows 
\begin{equation}
A^{(k)}_1 =A_{\text{topology}}^{(k)}+ A_{\text{phys}}^{(k)}
 \end{equation}
with $A_{\text{topology}}^{(k)}=A_{\text{edges}}^{(k)}+ A_{\text{tags}}^{(k)}$ and $A_{\text{phys}}= A_\iota$ referring to the physical quantum correlations among bulk intertwiners.

In the following analysis, we will consider the case where the intertwiners contribution is small, thereby looking at the correlations induced solely by the combinatorial structure of the graph, with a focus on the role of the tags. We leave the study of the intertwiners contribution in this setting for future work.

\section{Logarithmic Negativity and entanglement phases}\label{sct:phases}

By neglecting $A_\iota$, we are left with action $A_1^{(k)}$ with two-body interactions only, which favours neighbouring ``spins" to be parallel. Given $d_j=2j+1$ and $D_J=2J+1$ the dimension of edge spins and tag Hilbert space respectively, we define
\begin{equation}
    \log d_j\equiv\beta \quad,\qquad \log D_J\equiv\beta_t
\end{equation}
the interaction strengths intended as inverse temperatures, in analogy with the Ising model. Thereby, we can rewrite the action ${A}_{\text{topology}}^{(k)}$ as
\begin{align}
{A}_{\text{topology}}^{(k)}=\beta &\Biggr[\sum_{e^i_{vw} \in \mathcal{E}_R}\caldist{g_v}{g_w}+\sum_{e^i_{v\bar{v}} \in A_1}\caldist{g_v}{X}\nonumber\\ \nonumber&+\sum_{e^i_{v\bar{v}} \in A_2}\caldist{g_v}{X^{-1}}\Biggr]+\beta_t\sum_{v\in \mathcal{V}_R}\caldist{g_{v}}{\mathbbm{1}}\\
&= \beta \,H_e+\beta_t\, H_{\text{tags}}\label{topo-action}
\end{align}
with generalised hamiltonians $H_e, H_{\text{tags}}$ expressing the energy cost of the configuration. In general, we expect that the boundary conditions $X$, $X^{-1}$ and $\mathbb{1}$ percolate inside the network, flipping the generalized spins on the bulk vertices. This creates some internal domains filled with one of the three boundary conditions. Nonvanishing contributions to the action come from edges that connect different domains, which will be related to a non-zero Cayley distance. Given
\begin{equation} 
N_k(\rho_{A_1A_2})\simeq \sum_{\{g_v\}} e^{-A^{(k)}_{\text{topology}/0}}\, , \label{Partition2}
\end{equation}
in the large spin (the strong coupling or ``low temperature") regime, the leading contribution to $N_k$ corresponds to dominant $\{g_v\}$ configurations that minimize the action. Such configurations correspond to maximal uniform spin domains separated by extremal domain walls which minimize the energy cost (minimal surface areas) \cite{cepollaro2023}.  

Concretely, we then proceed minimizing the action in \eqref{topo-action} via a heuristic argument based on the property of the Cayley distance on the permutation group $S_k$. Given two permutations $g,h\in S_k$, the Cayley distance $\caldist{g}{h}$ defines a metric on the group, corresponding to the minimum number of swaps acting on $g$ to give $h$. For some special permutations, the Cayley distance depends nicely on the order on the group $k$. For example, in the case of the (anti-) cyclic permutations we deal with, we have
\begin{align}
    &\caldist{\mathbbm{1}}{X}=\caldist{\mathbbm{1}}{X^{-1}}=k-1\\&\caldist{X}{X^{-1}}=\begin{cases}k-1,\quad k\,\text{odd}\\k-2,\quad k\,\text{even}\end{cases}
\end{align}
With this metric, we can define a geodesic on the group $S_k$ between $g$ and $k$ as the set of permutations $\{\pi\}$ such that
\begin{equation}
    \caldist{g}{\pi}+\caldist{\pi}{h}=\caldist{g}{h}
\end{equation}
In particular, it is known that the set of permutations that are simultaneously geodesics between $X$, $X^{-1}$ and $\mathbb{1}$ exist, and it is in a bijection with the set of non crossing partitions $NC(k)$ of the set $\{1,\dots,k\}$, with cardinality given by the Catalan numbers $C_k$ \cite{Dong_21}. Defining by $\tau$ the permutations on such geodesics, we have that 
\begin{align}
    \caldist{\mathbb{1}}{\tau}=\left\lfloor\frac{k}{2}\right\rfloor,\quad\caldist{X}{\tau}=\caldist{X^{-1}}{\tau}=\left\lceil\frac{k}{2}\right\rceil-1
\end{align}
Given the homogeneity of our states, the minimization of the energy costs amounts to a minimization of the Cayley distances between any pairs of connected vertices. 

Let us consider a single vertex $v$. To such a vertex the random averaging assigns a permutation element $g_v$, which will eventually interact with the three permutations we choose as boundary conditions, namely $X$, $X^{-1}$ and $\mathbb{1}$. Recalling the symmetry between $X$ and $X^{-1}$, we expect that only \textit{three} possible \textit{minimal energy} configurations can result from such interaction:
\begin{itemize}
    \item $g_v=X,X^{-1}$, in which case we minimize the distance between the vertex $v$ and the neighboring vertices respectively contained in the $X$ or $X^{-1}$ domain;
    \item $g_v=\mathbb{1}$, in which case we minimize the distance between the vertex $v$ and the neighboring vertices contained in the $\mathbb{1}$ domain;
    \item $g_v=\tau$, in which we minimize simultaneously the distance between the $X$, $X^{-1}$ and the $\mathbb{1}$ domains.
\end{itemize}
Notice that the latter case is peculiar to the tripartite model only, since it can occur only when all three permutations interact in a single vertex \cite{cepollaro2023,Dong_21,Hayden:2016cfa}.

\subsection{Example: Cluster of four tags} \label{exesect}

\begin{figure}[t!]
\includegraphics[width=0.35\textwidth]{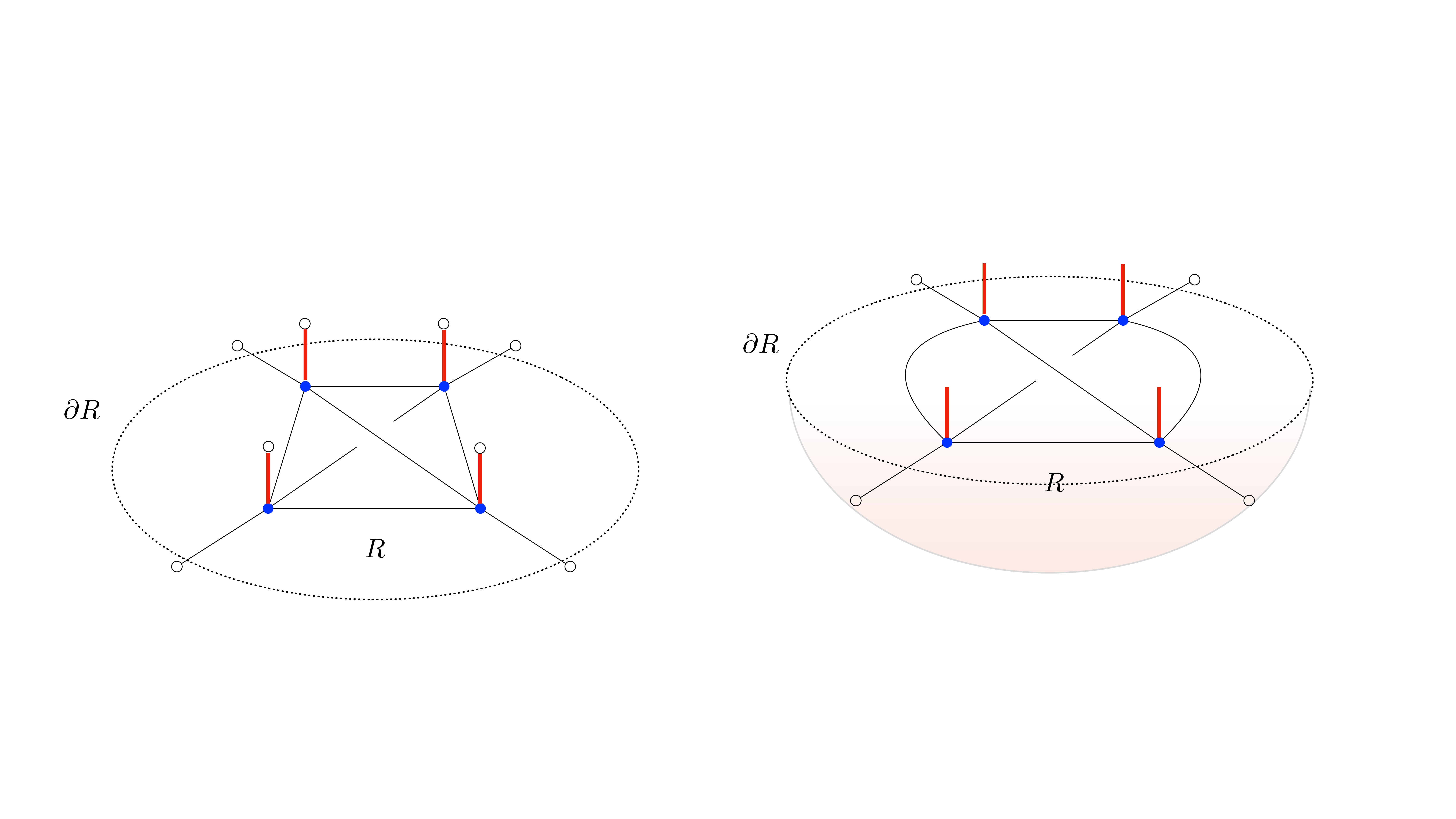}
\caption{Example: tagged open spin network graph corresponding to a cluster of four-valent tagged vertices, with $T=4$ and $E_{\partial R}=4$.}
\label{example}
\end{figure}

Let us consider the example of an open spin network state with support on a graph $\tilde{\gamma}_R$ with $E_{\partial R}=4$ boundary edges, and $V_R=4$ tagged 4-valent vertices in the bulk (see Fig.~\ref{example}). In the following we indicate with $T$ the number of tags, which in the present case is equal to the number of vertices, namely $T=V_R$. We refer to this setting as a \textit{cluster of tags}, corresponding to a uniformly curved  quantum $3D$ space region $R$. 
Via random averaging, we get four permutations operators on the spin network, one for each tagged vertex. Following the discussion above, we have three possible configurations with minimal energy. The corresponding actions (for even $k>0$) reads

\begin{enumerate}
    \item cluster colored with $X (X^{-1})$:
      \begin{equation}
      \mathcal{A}_k^{(X)}=\beta\, S\,(k-2)+\beta_t \,T\,(k-1)\,  
      \end{equation}
             Here $S$ is the \emph{minimal} number of edges cuts comprising the surface separating two domains with $X$ and $X^{-1}$. 
    \item cluster colored with $\mathbb{1}$:
      \begin{equation}\label{eq:enhole}
      \mathcal{A}_k^{(\mathbb{1})}=\beta E_{\partial R}\,(k-1)
      \end{equation}
    \item cluster colored with $\tau$:
      \begin{equation}
      \mathcal{A}_k^{(\tau)}=\beta E_{\partial R}\, \left(\frac{k}{2}-1\right)+\beta_t T\, \frac{k}{2}
      \end{equation}
\end{enumerate}

Such configurations define the three \textit{equilibrium phases} of our simple model, corresponding to three different entanglement regimes for the boundary state $\rho_{A_1A_2}$. The parameter space of such regimes is completely specified by the following three inequalities:
\begin{align}
&\mathcal{A}_k^{(\mathbb{1})}<\mathcal{A}_k^{(\tau)}\iff\beta E_{\partial R}<\beta_tT \label{cond1}\\[10pt]
&\mathcal{A}_k^{(\tau)}<\mathcal{A}_k^{(X)}\iff \beta E_{\partial R}<\beta_t T +2\beta S \label{cond2}\\[10pt]
&\mathcal{A}_k^{(X)}<\mathcal{A}_k^{(\mathbb{1})}\iff \beta E_{\partial R}>\beta_tT+\beta S\left(1-\frac{1}{k-1}\right)\label{cond3}
\end{align}
where the last condition \eqref{cond3} is relevant to establish a hierarchy between the three configurations, while it does not play a role in the search for the minimal energy cost. Notably, this rules out from the parameter space the only condition which depends explicitly on the order of the permutation group $k$.

\begin{figure}[t!]
\includegraphics[width=0.35\textwidth]{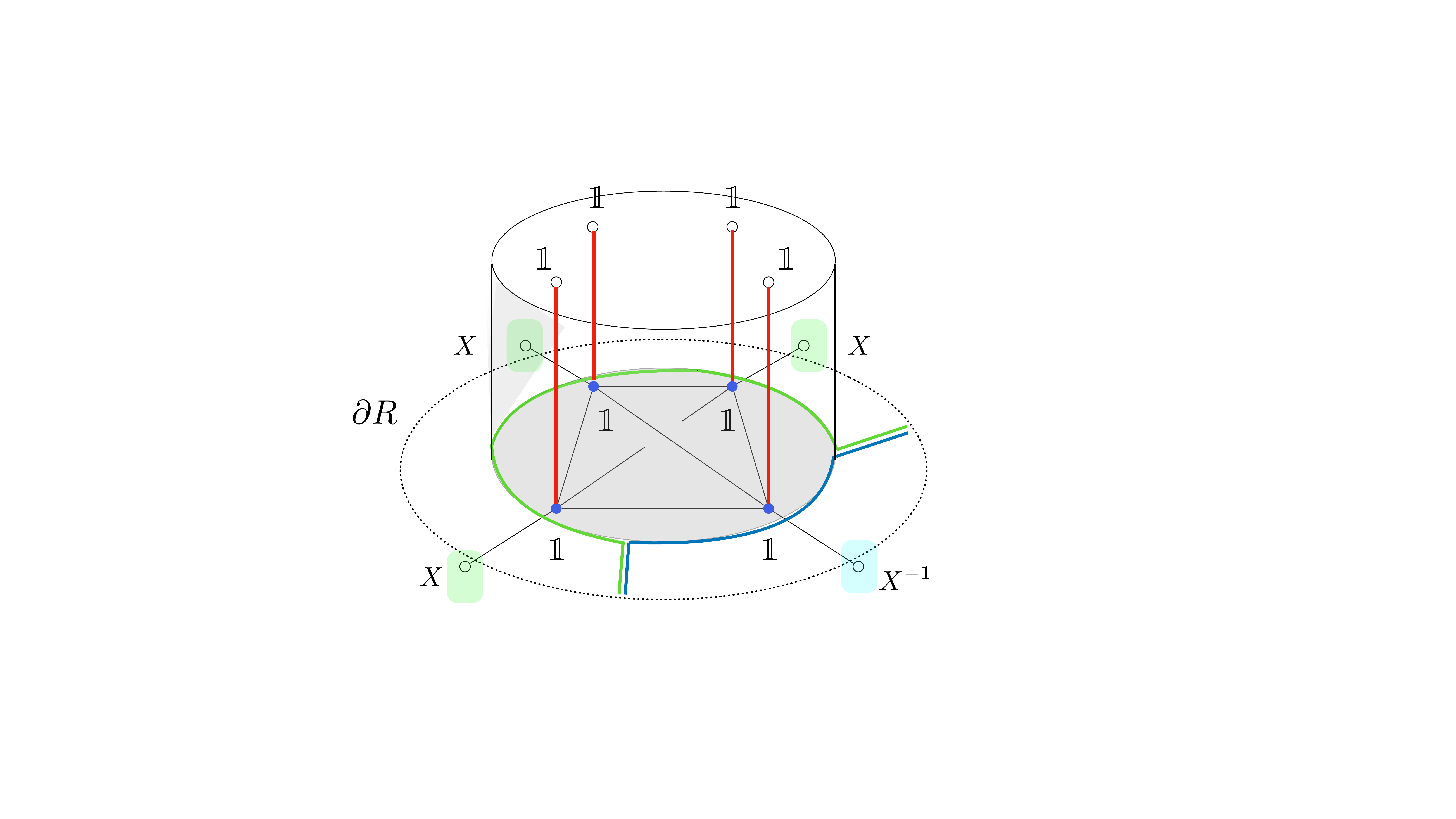}
\caption{Domain walls configuration in the hole regime.}
\label{holef}
\end{figure}

Consider the (positive definite) quantity $\beta E_{\partial R}$, namely the set of edges comprising the external surface of our $3D$ region weighted with the factor $\beta$. If $\beta E_{\partial R}<\beta_t T$, as $2\beta S>0$, also \eqref{cond2} is satisfied, then $\mathcal{A}_k^{(\mathbb{1})}<\mathcal{A}_k^{(X)},\mathcal{A}_k^{(\tau)}$. The configuration with a bulk colored with $g_v=\mathbb{1} \,\forall\, v$ dominates. In particular, this corresponds to the formation of a domain wall around the cluster of tags. The domain wall sharply divides the spin network boundary from the  bulk of the region $R$. We refer to such a regime as the \emph{hole} regime (see Fig.~\ref{holef}). 
Notice that, due to the bound on the maximum dimension of the tag spin \cite{Charles2016TheFS}, in the present example we can have at most
\begin{equation}
 J^{(max)}=4j\,\implies\,\beta_t^{(max)}=\beta+\log4\, .
\end{equation}
Then, in a large spins regime, we can generally consider $\beta_t^{(max)}\backsim\beta$ in facts. Thereby, in order for the \emph{hole} regime condition to be verified we must have 
\begin{equation}
    T > E_{\partial R}\, .
\end{equation}
More generally, the hole regime will be favoured by tags recoupling a large number of edge spins, so for coarse-grained vertices with high valence. 

\begin{figure}[t!]
\includegraphics[width=0.35\textwidth]{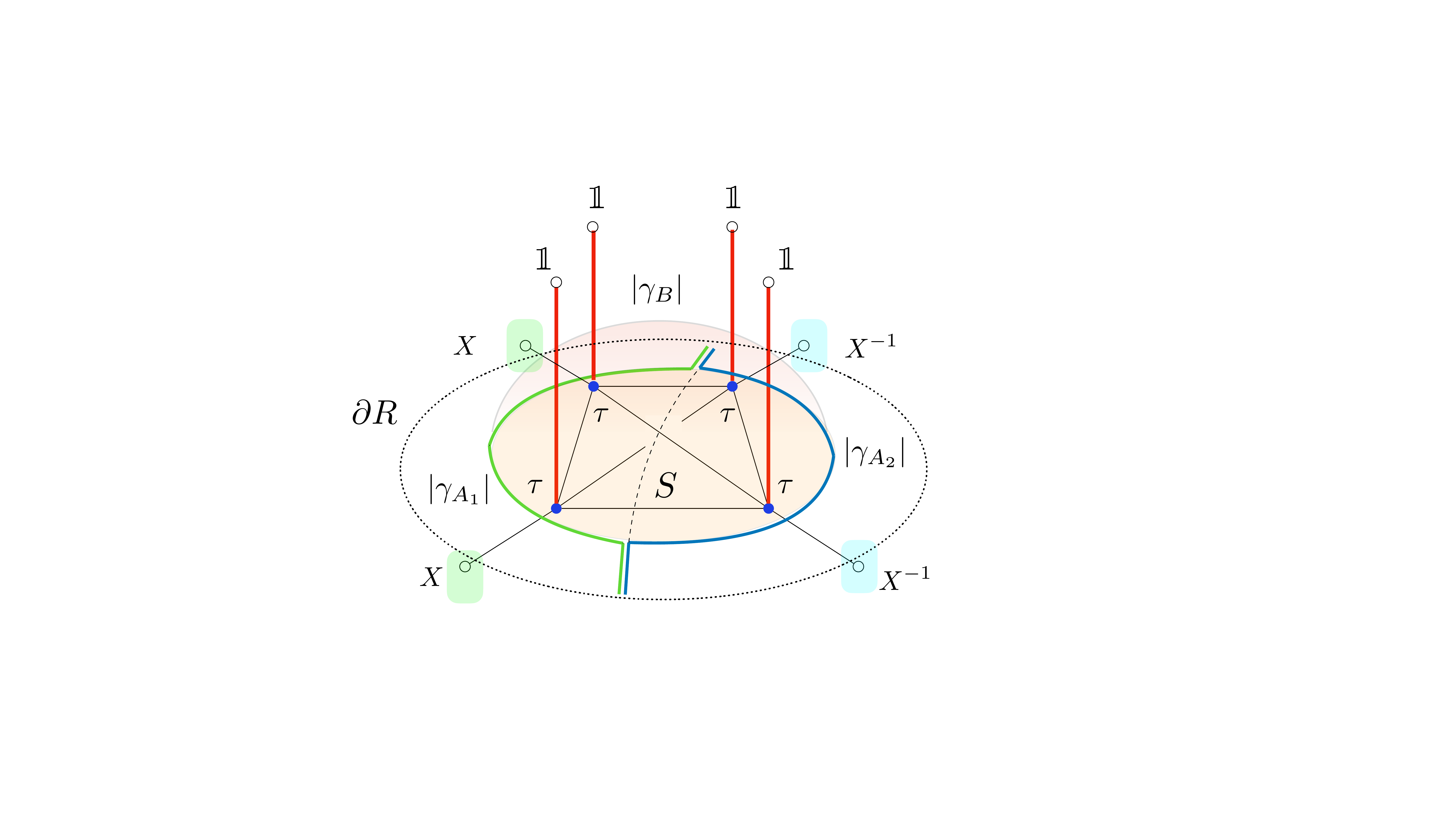}
\caption{Domain walls configuration in the island regime.}
\label{islandf}
\end{figure}

When the number of tagged vertices in the bulk and boundary edges is balanced we exit the hole regime and we need to consider $\beta E_{\partial R}\in[\beta_tT,\,\beta_tT+2\beta S]$. From \cref{cond1}, we see that in this range of values $\mathcal{A}_k^{(\tau)}<\mathcal{A}_k^{(\mathbb{1})}$. At the same time, \cref{cond2} implies that $\mathcal{A}_k^{(\tau)}<\mathcal{A}_k^{(X)}$ hence the minimal configuration is the one in which the geodesic permutation $\tau$ takes all the bulk, with the action $\mathcal{A}_k^{(\tau)}$ dominating. We call this setting the \emph{island} regime. The island domain wall stands between the three boundary conditions domains preventing them to access the bulk of $R$ (see Fig.~\ref{islandf}). Differently from the hole regime, the connectivity of the graph is not canceled in this case. Tags and boundary spins are connected through the island. 

\begin{figure}[t!]
\includegraphics[width=0.35\textwidth]{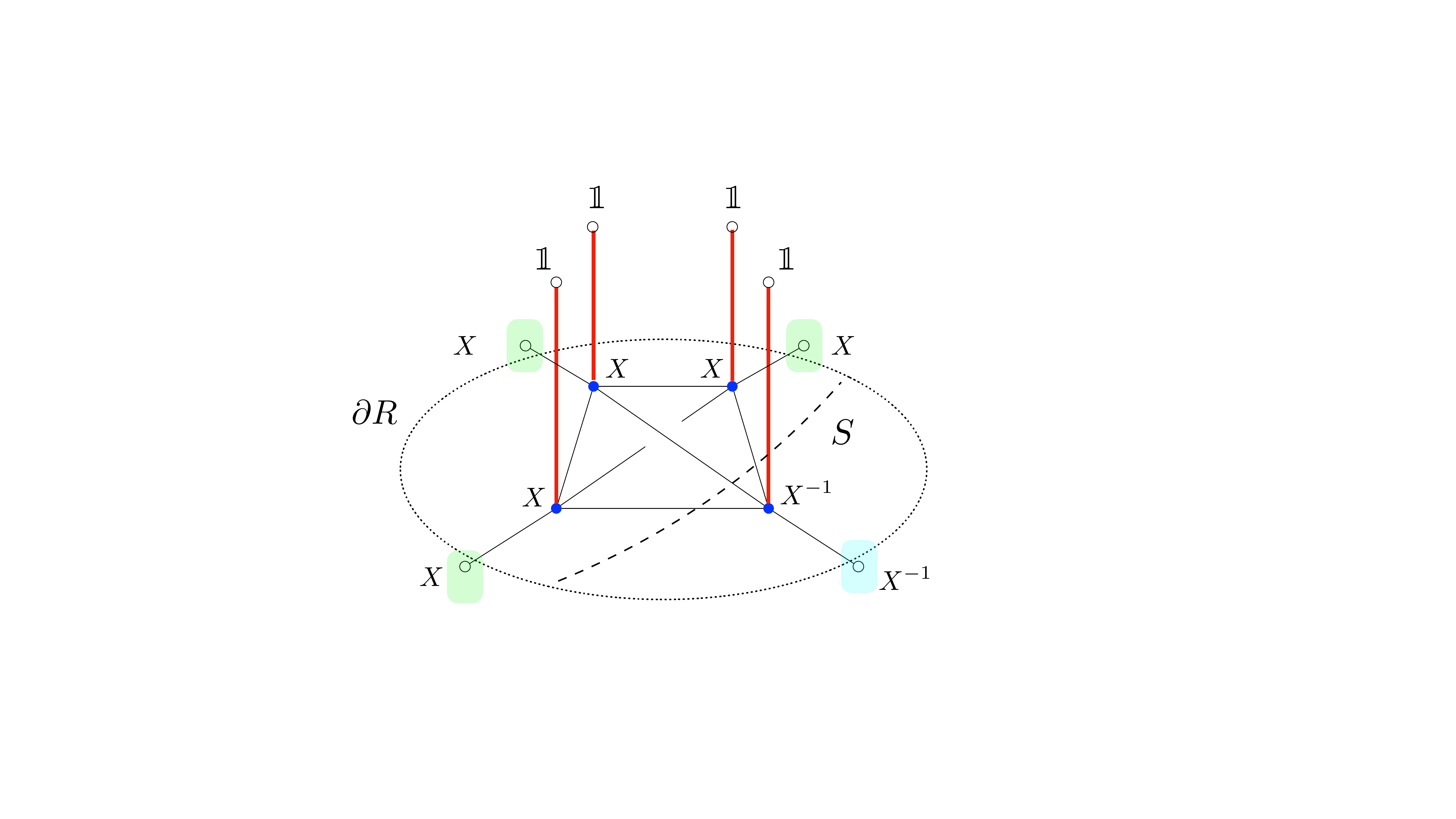}
\caption{Domain walls configuration in the bipartite regime.}
\label{bipartitef}
\end{figure}

Finally, for $\beta E_{\partial R}>\beta_t T+2\beta S$, namely when the tags environment is small compared with the boundary system, from \cref{cond1} and \cref{cond2} it follows that $\mathcal{A}_k^{(X)}<\mathcal{A}_k^{(\mathbb{1})},\mathcal{A}_k^{(\tau)}$. In this regime, the minimal energy configuration is associated to an effective bipartition of $R$ into two domains, one colored with the $X$ permutation and the other with the $X^{-1}$ permutation (see Fig.~\ref{bipartitef}).\footnote{The configurations in which on all the vertices in $R$ there is $X$ or $X^{-1}$ belong to this regime too. In these two extreme cases the domain wall $S$ is expelled from the bulk, and invades some of the boundary, in particular the region corresponding to the Hilbert space with minimal dimension among $\mathcal{H}_{A_1}$ and $\mathcal{H}_{A_2}$, which in turns is equivalent to consider the minimal area between the areas of the regions $A_1$ and $A_2$.} 
We refer to this setting as the \textit{bipartite} regime.

\subsection{Entanglement Phases}
We can now compute the typical logarithmic negativity of the random mixed state $\rho_{A_1A_2}$ in the three regimes considered in the given example. 

In the hole regime, from the action in \eqref{eq:enhole}, we get
\begin{multline}
     {E}_N^{(\text{hole})}(\rho_{A_1 A_2})\simeq-\lim_{k\to1}\mathcal{A}_{k}^{(\mathbb{1})}\\=-\lim_{k\to1}\beta E_{\partial R}(k-1)=0\, .
\end{multline}
A vanishing negativity tells us that the boundary state has Positive Partial Transpose (PPT), hence, according to the Peres criterion \cite{Peres_1996}, that boundary state could be separable. This suggests that $\rho_{A_1A_2}$ is well described by a maximally mixed state. The boundary system is almost fully entangled with the large number of degrees of freedom environment with little quantum correlation between $A_1$ and $A_2$. This effectively corresponds to a thermalization of the boundary state, in strong analogy with the results in~\cite{Anza:2017dkd, Shapourian_21}, where the typicality of the random state, in the large spin regime, similarly played a crucial role.

Moving to the island regime, the \emph{minimal} action reads
\begin{equation}
    \mathcal{A}_k^{(\tau)}=\beta E_{\partial R}\left(\frac{k}{2}-1\right)+\beta_t T\frac{k}{2} \label{islandaction}
\end{equation}
and the typical value of the logarithmic negativity in the $k\to1$ limit is given by
\begin{align}
    &E_N^{(\text{island})}(\rho_{A_1A_2})\simeq-\lim_{k\to1\nonumber}\left[\mathcal{A}_{k}^{(i)}+\log C_k\right]=\\&=-\lim_{k\to1}\left[\beta E_{\partial R}\left(\frac{k}{2}-1\right)+\beta_t T\frac{k}{2}+\log C_k\right]=\nonumber\\&=\frac{1}{2}(\beta E_{\partial R}-\beta_t T)+\log\frac{8}{3\pi} \, ,\label{Eisland}
\end{align}
with the Catalan numbers $C_k$ and $C_1=\frac{8}{3\pi}$ here taking into account the degeneracy in the choice of the geodesic coloring $\tau$ for the bulk. 

We have then a transition from an unentangled phase to an area law behavior, as expected by the topological character of the entanglement contributions. In fact, it is possible to write $E_N^{(\text{island})}(\rho_{A_1A_2})$ as a generalisation of the Ryu-Takayanagi formula for entanglement entropy \cite{Dong_21}. By defining the minimal surfaces that separate the entanglement wedges relative to each subsystem as $\abs{\gamma_{A_1}}$, $\abs{\gamma_{A_2}}$ and $\abs{\gamma_{B}}$, it is straightforward to see that 
\begin{equation}
    \abs{\gamma_{A_1}}= E_{A_1}\,,\quad\abs{\gamma_{A_2}}=E_{A_2}\,,\quad\abs{\gamma_{B}}=T
\end{equation}
where $E_{A_1}$, $E_{A_2}$ are the number of edges in $A_1$ and $A_2$ respectively.

Hence, since $E_{A_1}+E_{A_2}=E_{\partial R}$ we can write \cref{Eisland} as
\begin{multline}
    E_N^{(\text{island})}(\rho_{A_1A_2})=\frac{1}{2}\Bigr[\beta(\abs{\gamma_{A_1}}+\abs{\gamma_{A_2}})+\beta_t\abs{\gamma_B}\Bigr]\\+\log\frac{8}{3\pi}
\end{multline}
This is a generalisation of the RT formula  found in \cite{cepollaro2023} to the case of non-homogeneous mixed spin network states (two temperatures). As for the previous works, it can be expressed as a \textit{mutual information} between the subsystem $A_1$ and $A_2$:
\begin{align}
E_N^{(\text{island})}&=\frac{1}{2}\Bigr[S(\rho_{A_1})+S(\rho_{A_2})-S(\rho_{A_1A_2})\Bigr]+\log\frac{8}{3\pi}\nonumber\\
    &=\frac{1}{2}I(A_1:A_2)+\log\frac{8}{3\pi}
\end{align}

In the island regime, then, the logarithmic negativity is \emph{always} non-zero, and scales with the total area of the boundary of $R$, namely $\beta E_{\partial R}$, minus a negative correction that depends linearly on the number of tags. This implies that the boundary state is necessarily entangled. In particular, the presence of the environment reflects in a saturation of the entanglement of the two boundary subsystems. The logarithmic negativity only depends on the size of the system and of the environment but it is independent of the specific bipartition of the boundary.

 Finally, in the bipartite regime, the action with minimal energy is given by
\begin{equation}
    \mathcal{A}_k^{(X)}=\beta S(k-2)+\beta_t T(k-1) \, \label{bipartiteaction}
\end{equation}
and the typical logarithmic negativity reads
\begin{align}\label{Surf}
    &E_N^{(\text{bipartite})}(\rho_{A_1A_2})\simeq-\lim_{k\to1}\mathcal{A}_{k}^{(X)}=\\ \nonumber
    &=-\lim_{k\to1}\left[\beta S(k-2)+\beta_t T(k-1)\right]=\\ \nonumber
    &=\beta S
\end{align}
In this regime, negativity scales with the area of the internal domain wall $S$. 
In particular, we see from~\eqref{Surf} that in this regime we could achieve a PPT state for the boundary only by setting $S$ to 0, where a PPT state would correspond to a disconnected graph as expected. 

Such a behaviour is consistent with a limit toward a bipartite \emph{pure} boundary state setting corresponding to vanishing tags (trivial topology) \cite{Chirco:2021chk}.\\

Altogether, the effective topology of the bounded region is reflected in the multipartite entanglement of its \emph{tagged} spin-networks state description. Beside the toy model example, the proposed analysis can be easily generalised for graphs with generic number of vertices, with more than one open edge attached to each boundary vertex and with clusters of tags restricted to a subregion of the bulk graph. The universal character of the result is indeed associated to the typical behaviour of the randomised system in the large spin regime.

\section{Discussion} \label{sct:end}

In the framework of loop quantum gravity, quantum geometry states corresponding to \emph{bounded} regions of $3D$ space with \emph{uniform curvature} can be synthetically described via superpositions of \emph{tagged} spin-networks with boundary.
Curvature naturally builds up at the spin-network vertices in the form of topological defects described by spin tags, as the result of the \emph{partial tracing} over nontrivial $SU(2)$ holonomies of the Ashtekar connection along the network.

In presence of a boundary, a quantum geometry wave-function generally behaves as a map encoding the bulk information into the boundary Hilbert space. In the proposed analysis, we generalize such a bulk-to-boundary mapping  as to include the space of bulk topological defects, effectively described as extra (inner) boundary degrees of freedom.\footnote{In fact, we can think of bulk tags as the recoupled dangling bulk spins dual to the surface of holes cut out of planar spin network graph.}
In the resulting \emph{extended boundary} state, bulk information is \emph{shared} among generically entangled boundary spins and tags: the surface and the intrinsic curvature of the quantum space region are entangled. Remarkably, the degree of quantum correlations among boundary subregions is affected by the presence of the bulk curvature, which plays the role of a hidden environment for the  mixed outer boundary state. 

The effect of the curvature on the boundary entanglement is the focus of our analysis. We model the generalised boundary mapping on a tripartite ($A_1,A_2, B$) system and consider the reduced boundary state $\rho_{A_1A_2}$ obtained via a trace over the space of tags (system $B$). We further characterise the coarse viewpoint of a boundary observer by considering a \emph{random measurement} on the bulk state, thereby ultimately dealing with a \emph{random mixed state} for the boundary.

The entanglement of the bipartite mixed boundary is quantified via a measure of logarithmic negativity, along the lines of the recent results in~\cite{Dong_21,Shapourian_21,cepollaro2023}. 
Concretely, we restrict the analysis to a class of states characterised by graphs coloured by the same spin $j$ on each edge and dressed with a tag of corresponding recoupled spin $J$ at each vertex. We call this setting a \emph{cluster of tags}. The full computation is carried on for a simple cluster of four tags is given in Section~\ref{exesect}. By using standard replica techniques for random tensor network, this computation is ultimately mapped to the evaluation of the minimal action of a classical generalised Ising-like statistical model on the spin network graph, in the limit of vanishing temperature (see~\cite{cepollaro2023} and references therein).

As a first general result, we find that the Ising-like model action dual to the random spin network generally splits into two main contributions:
\begin{equation}\label{twoa2}
A^{(k)}_1 =A_{\text{topology}}^{(k)}+ A_{\text{phys}}^{(k)}\,
 \end{equation} 
that is a contribution $A_{\text{topology}}^{(k)}$ which accounts for the entanglement induced by the \emph{topology} of the spin network graph and a second one, $A_{\text{phys}}^{(k)}$, encoding the bulk intertwiner entanglement. The former, in particular, further splits into two well distinct contributions, 
$$A_{\text{topology}}^{(k)}=A_{\text{edges}}^{(k)}+ A_{\text{tags}}^{(k)}\, ,$$ 
respectively associated to the \emph{connectivity} of the bulk network (trivial topology), induced by maximally entangled edge states, and to the presence of the tags. The last term is therefore specific to the case of spin network states with topological defects. 

In a topologically trivial setting, as showed in \cite{cepollaro2023}, the first term in \eqref{twoa2} is associated to an area law scaling of the entanglement of the boundary, while bulk intertwiners entanglement can be interpreted in analogy with \cite{Hayden:2016cfa}.
We leave the characterisation of the random bulk contribution for future work, while focusing on the nontrivial topological contribution to the action.

Within the typical regime, the entanglement phases of the random boundary state can be described solely in terms of the relative dimensions of the three subsystems 
$A_1$, $A_2$ and $B$,\footnote{Where $\text{dim}(A_1)=(d_j)^{E_{A_1}}$, $\text{dim}(A_2)=(d_j)^{E_{A_2}}$ and $\text{dim}(B)=(d_J)^{T}$ respectively.} via the parameter
\begin{equation}
q=\beta_t T/ \beta E_{\partial R} \label{qqq} = \frac{\log[\text{dim}(B)]}{\log[\text{dim}(A_1) \text{dim}(A_2)]}\, ,
\end{equation}
which expresses the ratio of bulk curvature over boundary surface, with $E_{\partial R}= E_{A_1}+E_{A_2}$. 
Two main entanglement phases are separated by a critical point at $q=1$.

For $1-2 S/E_{\partial R}<q<1$, the curvature environment mediates the entanglement of the boundary. The logarithmic negativity, modulo degeneracy, scales with the area of the cluster boundary with a negative correction which depends
linearly on the number of tags. We have
\begin{align}  \label{mutual}
    E_N^{(\text{island})}(\rho_{A_1A_2})&\propto \frac{1}{2}\beta E_{\partial R}(1-q)\\ \nonumber 
    &=\frac{1}{2}\Bigr[\beta(\abs{\gamma_{A_1}}+\abs{\gamma_{A_2}})+\beta_t\abs{\gamma_B}\Bigr]
\end{align}
with $\abs{\gamma_{A_1}}= E_{A_1}$ ,$\abs{\gamma_{A_2}}=E_{A_2}$ and $\abs{\gamma_{B}}=T$, the minimal surfaces separating the entanglement wedges relative to each subsystem. This is a generalisation of the Ryu-Takayanagi formula for entanglement entropy for the tripartite setting, in agreement with \cite{cepollaro2023, Dong_21} and it can be expressed as a \textit{mutual information} between the subsystem $A_1$ and $A_2$.

\begin{figure}[t!]
\includegraphics[width=0.45\textwidth]{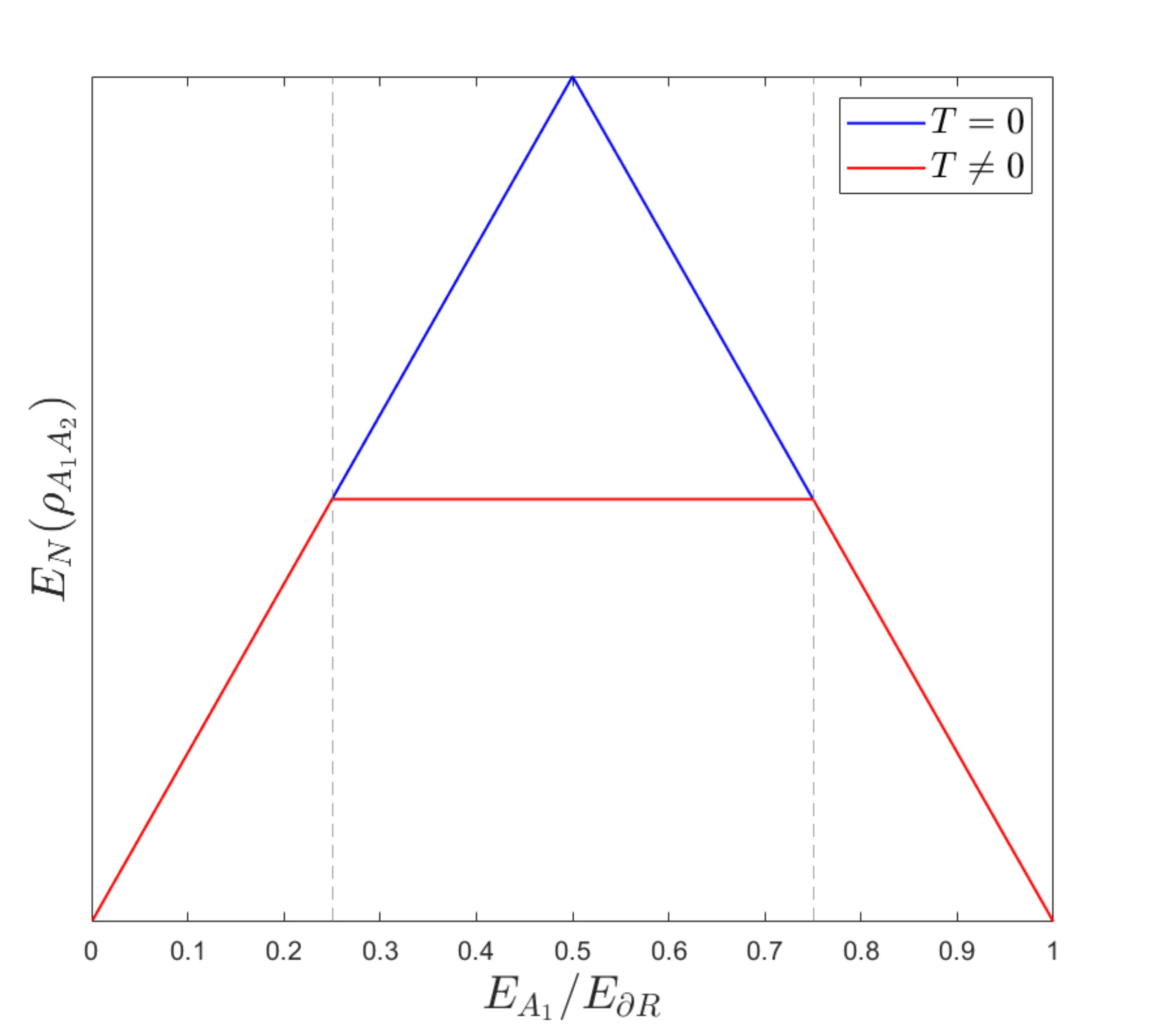}
\caption{In red the generalized Page curve of the logarithmic negativity for the random mixed state $\rho_{A_1A_2}$. In red, how the Page curve would appear for random pure states.}
\label{pagef}
\end{figure}

Finally, for  $q<1-2 S/E_{\partial R}$, when the curvature environment is much smaller than the boundary system, we tend to a bipartite setting and the typical logarithmic negativity  scales with the area of the surface $S$ separating $A_1$ and $A_2$ through the boundary. In this case, we find
\begin{equation}
E_N^{(\text{bipartite})}(\rho_{A_1A_2})\propto \beta S\, .
\end{equation}

Notice that the value of $S$ cannot be tuned in our analysis, as the surface is fixed by the spin network connectivity for the given graph. However, assuming some regularity of the graph, one can geometrically relate this value to the  bipartition of the boundary. For instance, assuming a generic tagged graph with a large density of vertices, dual to a 3-ball, we can imagine the minimal surface $S$ to be approximated by the intersection of the sphere and a generic plane, which divides $A_1$ from $A_2$ on the boundary of $R$. In this case, we can approximately set
\begin{equation}
S\approx j^2\frac{E_{A_1}E_{A_2}}{E_{\partial R}}
\label{circlearea}
\end{equation}
Therefore, we see that the island condition is favoured when the dimension of the two subsystems $A_1$ and $A_2$ is well balanced, since $S$ is maximized by high values of both $E_{A_1}$ and $E_{A_2}=E_{\partial R}-E_{A_1}$. Conversely, from the condition $q<1-2 S/E_{\partial R}$, the bipartite regime is favoured when the two boundary subregions $A_1$ and $A_2$ are extremely asymmetric in extension (smaller $S$). 

In particular, we can describe the transition from the island to the bipartite regime by using \eqref{circlearea} in the condition $\mathcal{A}_k^{(bipartite)}<\mathcal{A}_k^{(island)}$. We find
\begin{equation}
\left(\frac{E_{A_1}}{E_{\partial R}}\right)^2-\frac{E_{A_1}}{E_{\partial R}}+\left(1-q\right)> 0
\end{equation}
Thereby, for $0<q< 1$, we can identify two critical points between the island and the bipartition phases,
\begin{equation}
    c^*_{1,2}(T)=\frac{1}{2}\pm\frac{1}{2}\sqrt{q}\, .
\end{equation}
The behaviour of the negativity is well described by the generalized Page curve in \cref{pagef}.
As a general feature of a tripartite random state, we find that the degree of correlation of the two boundary subregions $A_1, A_2$ depends on the degree of quantum correlations the two boundary systems have with the curvature environment. By varying the relative size of the two subsystems, the logarithmic negativity shows an initial increase and a final decrease in analogy with the Page curve. When $T=0$ (trivial topology), $c^*_1(0)=c^*_2(0)=1/2$ and we get back the standard Page curve \cite{PhysRevLett.71.1291} (no tripartition). As soon as $T\neq 0$, the intermediate island regime forms and the negativity has a plateau. In this regime the logarithmic negativity only depends on the size of the system and of the environment but not on how the system is partitioned.

On the other hand, for $q>1$, corresponding to a large tags environment with respect to the boundary system, the logarithmic negativity vanishes. The boundary system is almost fully entangled with the environment with little quantum correlation between $A_1$ and $A_2$. In particular, the area law behaviour is lost. This suggests that large curvature is associated to a \emph{maximally} mixed boundary state $\rho_{A_1A_2}$. Such a thermalization of the boundary due to curvature is in strong analogy to the quantum gravity formulation of the hoop conjecture proposed in ~\cite{Anza:2017dkd}, where again the typicality of the random state, in the large spin regime, plays a crucial role.

If we consider that the large spin regime we have been working in is necessarily associated with a semiclassical limit for our quantum geometry states, then we can tentatively interpret the formulas found for the entanglement negativity in terms of areas of a bounded $3D$ region of a Riemannian manifold and look for a direct relation with its Ricci curvature. Indeed, we know that the value of the scalar curvature $\mathcal{R}$ of a Riemannian $n$-manifold $\mathcal{M}$ at a point $p$  can be quantified by the ratio of the area of the $(n-1)$-dimensional boundary of a ball of radius $\varepsilon$ in $\mathcal{M}$ to that of a corresponding ball in the Euclidean space. For small $\varepsilon$, the ratio is given by~\cite{gallot2004riemannian, Chavel1984}
\begin{equation}\label{ratio}
\frac{\text{Area}(\partial B_{\varepsilon}(p) \subset \mathcal{M})}{\text{Area}(\partial B_{\varepsilon}(0) \subset \mathbb{R}^n)}= 1-\frac{\mathcal{R}}{6n}\varepsilon^2+ O(\varepsilon^3)\, .
\end{equation}
Now, in the semiclassical limit, for $n=3$, we can easily associate the area of the boundary of a flat three ball region with the sum of the areas of the triangles dual to the open edges spins comprising the boundary of $R$ (no tags), that is $\text{Area}(\partial B_{\varepsilon}(0)) \simeq\beta E_{\partial R}(\varepsilon)$, where $E_{\partial R}(\varepsilon)$ counts the number of boundary edges at a certain radius $\varepsilon$, now measured via some distance on the spin network graph. In case of the curved three ball in $\mathcal{M}$, however, we expect this value to be modified by the presence of tags. In particular, if we assume  \begin{equation}
\text{Area}(\partial B_{\varepsilon}(p))\simeq \beta E_{\partial R}(\varepsilon) +\kappa\, \beta_t T(\varepsilon)\, 
\end{equation}
namely a linear correction in the number of tags (positive or negative depending on the sign of $\kappa$), where $T(\varepsilon)$ indicates the number of tags within a certain radius $\varepsilon$. Thereby, the ratio in \eqref{ratio} gives us a direct relation
\begin{equation}\label{ratio2}
q(\varepsilon) \simeq - \frac{\mathcal{R}}{18\, \kappa}\varepsilon^2 
\end{equation}
between the entanglement phase parameter and the Ricci curvature. For instance, in this light, we find a nice relation between the area scaling negativity (or mutual information) in \eqref{mutual} and the geometric relation in \eqref{ratio}. Such a relation is consistent with the interpretation of our results and hints toward a characterisation of the curvature in purely information theoretic terms \cite{noi}. 

Let us then conclude with a few further remarks.
A first remark is in order concerning the non-gauge invariant character of the degrees of freedom considered in our analysis. Spin network's tags are not different from boundary dangling spins, whose presence indeed breaks gauge invariance at the boundary. The entanglement we describe among boundary spins and tags cannot be measured in terms of gauge invariant observables, hence it is considered as non physical in loop quantum gravity~\cite{Rovelli:2013fga}. Nevertheless, we expect such a topological layer of entanglement to become essential in modelling the emergence of spacetime connectivity as soon as we allow both edge and tags spins to be physical at some higher energy scale, where we expect gauge invariance to break. This perspective would require an extension of the kinematical description of quantum geometry in loop quantum gravity, which is in part provided by the second quantized formalism of group field theory~\cite{Oriti:2013aqa}. Differently, one can take the present analysis as a toy model for an ultimate description of the entanglement structure of \emph{edge modes} degrees of freedom at the interface of subsystems~\cite{Donnelly:2020xgu,Ciambelli:2021nmv, Kabel:2022efn}. In this sense, we expect the described universal features of the interface entanglement in the large spin regime to be general. 

\begin{figure}
\includegraphics[width=0.42 \textwidth]{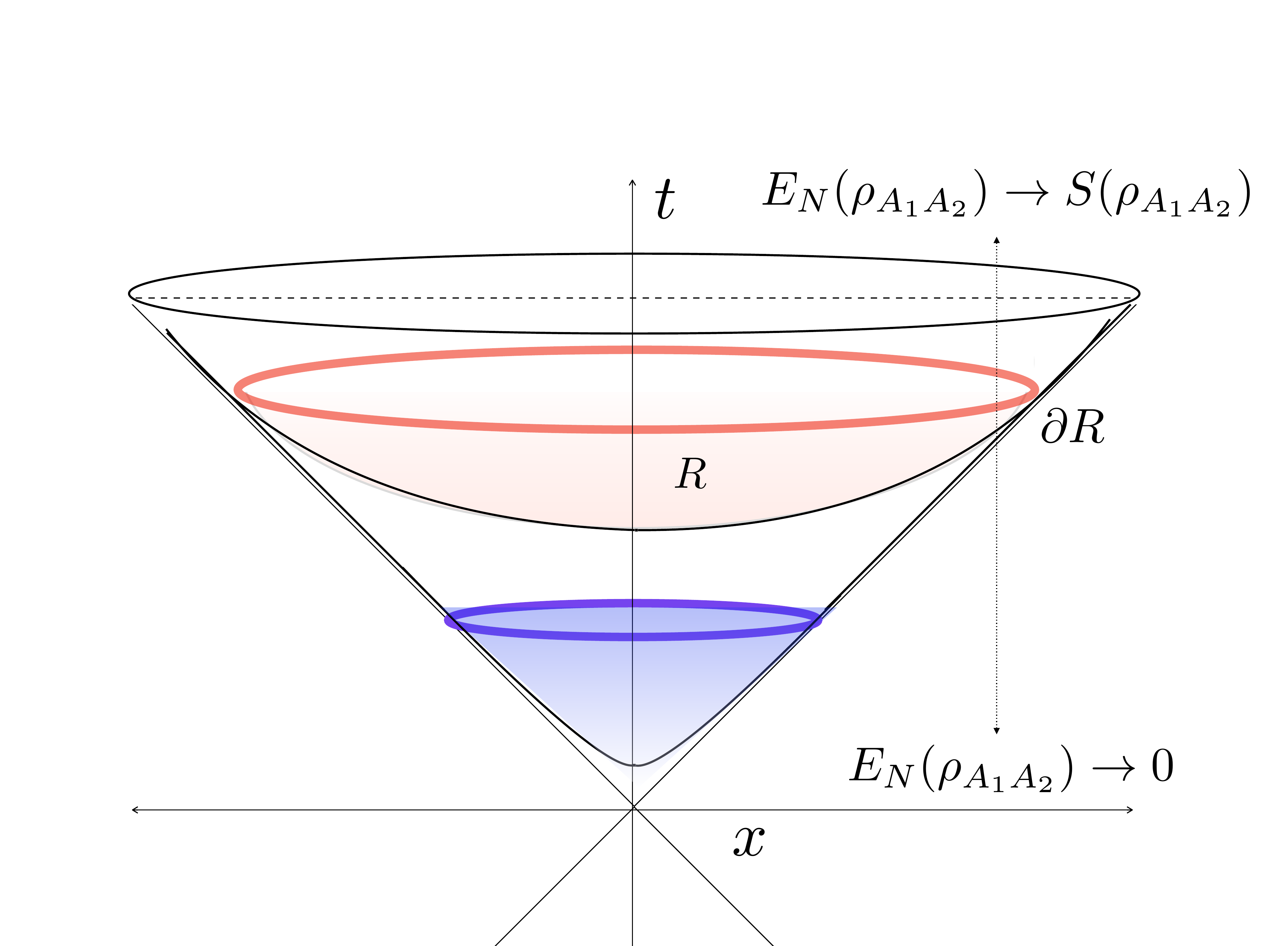}
\caption{Milne spacetime sliced by $3D$ surfaces with hyperbolic embedding. Close to Milne's big bang, the quantum realisation of such surfaces have vanishing negativity, $E_N(\rho_{A_1A_2})\to 0$. As time increases, the negativity of the $2D$ cosmological horizon grows and tends to bipartite entanglement entropy,  $E_N(\rho_{A_1A_2})\to S(\rho_{A_1A_2})$ at large times.}
\label{milne}
\end{figure}

A final remark has to do with the interplay of intrinsic and extrinsic geometry in loop quantum gravity. As showed in \cite{Livine:2019cvi, Freidel:2018pvm}, in loop quantum geometry states, non-trivial Ashtekar-Barbero holonomies can be thought alternatively as intrinsic geometry or \emph{extrinsic curvature}. In particular, it is possible to compensate the defects of the intrinsic geometry appearing while coarse-graining spin networks by Lorentz boosts which change the local embedding of the spin network description of quantum $3D$ space in the $(3+1)$-D geometry. This  suggests a natural relation between embedding and entanglement to which we shall dedicate further investigation in future work. For instance, one could figure the phase transition previously described, between a maximally entangled $(q\ll1)$ to a maximally mixed $(q\gg1)$ boundary state, as being described by a sequence of quantum $3D$ spacelike slices with a gradient in curvature. If we embed such slices, via boost, in the hyperbolic space, we would get a transition from  almost flat bounded regions with large corner surface corresponding to a maximally entangled boundary state to highly curved regions with small corner surface corresponding to maximally mixed boundary state. A similar setting naively recalls the geometry of the \emph{Milne spacetime} cosmology~\cite{Mukhanov:2005sc} (see Fig.~\ref{milne}). Intriguingly, in this analogy, the proposed analysis would suggest that, close to the big bang, quantum $2D$ boundary surfaces would be described by maximally mixed (or thermal) states, while as the universe expands and flattens, quantum correlations between subregions of the surface would increase. An accurate analysis of a similar scenario is left for future work.

\begin{acknowledgments}
The authors would like to thank Alioscia Hamma and Daniele Oriti for useful discussions on the preliminary results of the work. G.C. would like to thank in particular Michele Arzano for suggesting the relation between a set of boost embedded quantum 3D space regions and Milne geometry.
\end{acknowledgments}

\bibliography{bibliography_neg.bib}
\bibliographystyle{apsrev4-1}
\end{document}